\documentclass[12pt]{article}

\usepackage[top=1in, bottom=1in, left=1in, right=1in]{geometry}

\usepackage{pdflscape} 

\usepackage{hyperref}

\usepackage{titlesec}
\titlespacing\section{0pt}{12pt}{0pt}
\titlespacing\subsection{0pt}{6pt}{0pt}
\titlespacing\subsubsection{0pt}{6pt}{0pt}

\usepackage[T1]{fontenc}
\usepackage{textcomp}
\usepackage{listings}
\usepackage{xcolor}
\usepackage{tcolorbox}

\tcbuselibrary{listings,breakable}

\newtcblisting{codeblock}{
  listing only,
  breakable,
  colback=gray!5,
  colframe=gray!40,
  boxrule=0.3pt,
  arc=2pt,
  left=6pt,
  right=6pt,
  top=6pt,
  bottom=6pt,
  listing options={
    basicstyle=\ttfamily\small,
    breaklines=true,
    columns=fullflexible,
    escapeinside={@}{@}
  }
}
\usepackage{titlesec}
\usepackage{CJKutf8}

\definecolor{cornflowerblue}{RGB}{100,149,237}

\newtcblisting{codeblocks}{
  colback=white,
  colframe=cornflowerblue,
  listing engine=minted,
  listing only,
  minted language=text,
  minted options={
    escapeinside=@@,
    fontsize=\small,
    breaklines=true,
    breaksymbolleft=\phantom{→} 
  },
  boxrule=0.3mm,
  sharp corners,
  left=0mm,
  right=0mm,
  top=0.5mm,
  bottom=0.5mm,
}
\definecolor{salmon}{rgb}{0.980, 0.502, 0.447}

\newtcblisting{codeblockv}{
  colback=white,
  colframe=salmon,
  listing engine=minted,
  listing only,
  minted language=text,
  minted options={
    escapeinside=@@,
    fontsize=\small,
    breaklines=true,
    breaksymbolleft=\phantom{→}
  },
  boxrule=0.3mm,
  sharp corners,
  left=0mm,
  right=0mm,
  top=0.5mm,
  bottom=0.5mm,
}

\usepackage{titletoc}
\usepackage{lipsum}

\usepackage{lineno} 
\usepackage{times} 
\usepackage[table]{xcolor}
\usepackage[normalem]{ulem}
\usepackage{longtable,booktabs} 
\usepackage{footnote}
\usepackage{tablefootnote} 
\usepackage{capt-of}
\usepackage{booktabs}
\usepackage{amsmath}
\usepackage{amsthm}
\usepackage{amssymb}
\usepackage{bbold} 
\usepackage{listings}
\usepackage{multicol,multirow,makecell}

\usepackage{url}
\usepackage{threeparttable}
\usepackage{lscape}
\usepackage{longtable,booktabs} 
\usepackage{geometry}
\usepackage{verbatim}
\usepackage{enumitem}
\usepackage{titlesec}
\usepackage{tabularx}
\usepackage{tikz}
\usepackage{subcaption}
\usepackage{graphicx}
\usepackage{float}
\usepackage[bottom]{footmisc}

\usepackage{CJK} 
\usepackage{hhline} 

\usepackage{natbib}

\usepackage{multibib}
\newcites{app}{References}%
\usepackage{ulem}
\usepackage{soul}
\usepackage{tabularray}
\UseTblrLibrary{booktabs}

\usepackage{setspace}
\usepackage{appendix}

\usepackage{setspace}
\usepackage{placeins}
\usepackage{authblk}
\titlespacing\section{0pt}{10pt}{10pt} 
\titlespacing\subsection{0pt}{8pt}{8pt} 

\hypersetup{
	colorlinks   = true, 
	urlcolor     = blue, 
	linkcolor    = blue, 
	citecolor   = blue    
}

\usepackage{hyperref}

\usepackage{array}
\newcommand{\PreserveBackslash}[1]{\let\temp=\\#1\let\\=\temp}
\newcolumntype{C}[1]{>{\PreserveBackslash\centering}p{#1}}
\newcolumntype{R}[1]{>{\PreserveBackslash\raggedleft}p{#1}}
\newcolumntype{L}[1]{>{\PreserveBackslash\raggedright}p{#1}}

\definecolor{mypink1}{RGB}{255, 204, 255}
\definecolor{mypurple1}{RGB}{204,51,255}
\definecolor{mygreen1}{RGB}{61,245,0}
\definecolor{myorange1}{RGB}{255,102,0}
\definecolor{cornflowerblue}{RGB}{100,149,237}
\definecolor{salmon1}{RGB}{250,128,114}
\newcommand*{\rom}[1]{\expandafter\@slowromancap\romannumeral #1@}

\newcommand{\hide}[1]{} 

\makeatletter
\newcommand{\samethanksmark}{%
  \textsuperscript{\textcolor{\@linkcolor}{*}}%
}
\makeatother

\theoremstyle{plain}

\theoremstyle{definition}

\newcommand{\RNum}[1]{\uppercase\expandafter{\romannumeral #1\relax}}

\begin{document}

\title{
Understanding the Mechanism of Altruism \\ in Large Language Models
}

\author[1]{Shuhuai Zhang}
\author[2]{Shu Wang}
\author[3]{Zijun Yao}
\author[4]{Chuanhao Li}
\author[5\thanks{Corresponding authors}]{Xiaozhi Wang}
\author[6\samethanksmark]{Songfa Zhong}
\author[7\samethanksmark]{Tracy Xiao Liu}

\affil[1]{\small School of Economics, Central University of Finance and Economics}
\affil[2]{\small Lingnan College, Sun Yat-sen University}
\affil[3]{\small Department of Computer Science and Technology, Tsinghua University}
\affil[4]{\small Department of Industrial Engineering, Tsinghua University}
\affil[5]{\small Shenzhen International Graduate School, Tsinghua University}
\affil[6]{\small Department of Economics, Hong Kong University of Science and Technology}
\affil[7]{\small School of Economics and Management, Tsinghua University}

\date{\today}

\maketitle

\vspace{-1.5cm}

\begin{abstract}
Altruism is fundamental to human societies, fostering cooperation and social cohesion. While recent studies suggest that large language models (LLMs) can display human-like prosocial behavior, the underlying mechanism that produces such behavior remains unclear. We leverage sparse autoencoders (SAEs) to identify a set of features that contribute to an LLM's altruistic choices in the Dictator Game. We show that intervening on these features substantially shifts the model's allocations and produces similar effects in several other social decision scenarios. Together, these results link the model's prosocial output to internal features that can be identified and causally manipulated, providing a transparent approach to studying the computational basis of decision-making in LLMs.   
\end{abstract}

\medskip

\noindent\textit{Keywords}: Large Language Models (LLMs), Mechanistic Interpretability, Sparse Autoencoders, Altruism

\medskip

\noindent\textit{JEL Classification Numbers}: C9, C45 

\clearpage

\section{Introduction}
Altruism shapes individual well-being, community cooperation, and social functioning \citep{trivers1971evolution,axelrod1981evolution,fehr2003nature,nowak2006five,Awad2018,binz2025foundation}. A substantial body of research across biology, neuroscience, psychology, and economics has clarified altruism's role in human evolution, its neural mechanisms, as well as its behavioral determinants and consequences \citep{brosnan2003monkeys,harbaugh2007neural,henrich2010markets,tricomi2010neural,fisman2015distributional,hsu2015}. In parallel, large language models (LLMs) have rapidly evolved into decision-support tools and are increasingly deployed as autonomous agents in social and economic settings \citep{brynjolfsson2017can,brinkmann2023machine,noy2023experimental,burton2024large,brynjolfsson2025generative,kobis2025delegation}. Existing behavioral studies have reported that LLMs can produce prosocial output in standard economic games, sometimes at levels comparable to humans \citep{chen2023emergence,mei2024turing,akata2025playing,johnson2025testing,xie2025using}. 
However, the computational mechanisms that generate such prosocial outputs remain unclear.

This gap matters because observing prosocial behavior alone does not reveal how such behavior is generated. For example, a model that makes an altruistic choice in a particular economic game may do so because it has learned a generalizable pattern for social decision-making, or simply because it responds to context-specific cues in that particular setting. Without understanding the underlying mechanisms, it is unclear what drives these behaviors, how these behaviors can be reliably modified, and whether the same mechanisms will produce similar behavior in other contexts.

Recent advances in mechanistic interpretability provide tools for studying the internal processes of LLMs. When an LLM reads a prompt and generates a response, it produces a high-dimensional pattern of internal activity. Many different signals are mixed together in that activity. Sparse autoencoders (SAEs) offer a promising way to make these patterns more tractable. They express the model's internal representation as a combination of \textit{sparse features}. These sparse features are easier to analyze and interpret, as they often correspond to recognizable meanings. Observing how these features are expressed during response generation can help reveal which concepts or processes may be involved in producing particular outputs \citep{elhage2022superposition,bricken2023monosemanticity}. Building on work that uncovers structured representations in language models \citep{gurnee2023language, templeton2024scaling, chen2025financial,movva2025sparse}, we identify features associated with altruistic behavior and, importantly, test whether manipulating these features causally changes the model’s outputs.

We conduct our analysis on Llama-3.1-8B-Instruct \citep{grattafiori2024llama}, which reads the prompts used in our tasks and generates a response. To examine the model’s internal activity, we use OpenSAE \citep{opensae}, an open-source toolkit that provides pretrained sparse autoencoders for the Llama-3.1-8B model family. OpenSAE maps Llama’s internal representation onto interpretable SAE features and allows us to intervene on selected features by increasing or decreasing their activity  while the model generates a response. We use this combination because Llama is a widely used open-weight model, and OpenSAE provides the pretrained components needed for feature-level measurement and intervention. 

\begin{figure*}[t!]
\centering
\includegraphics[width=\textwidth]{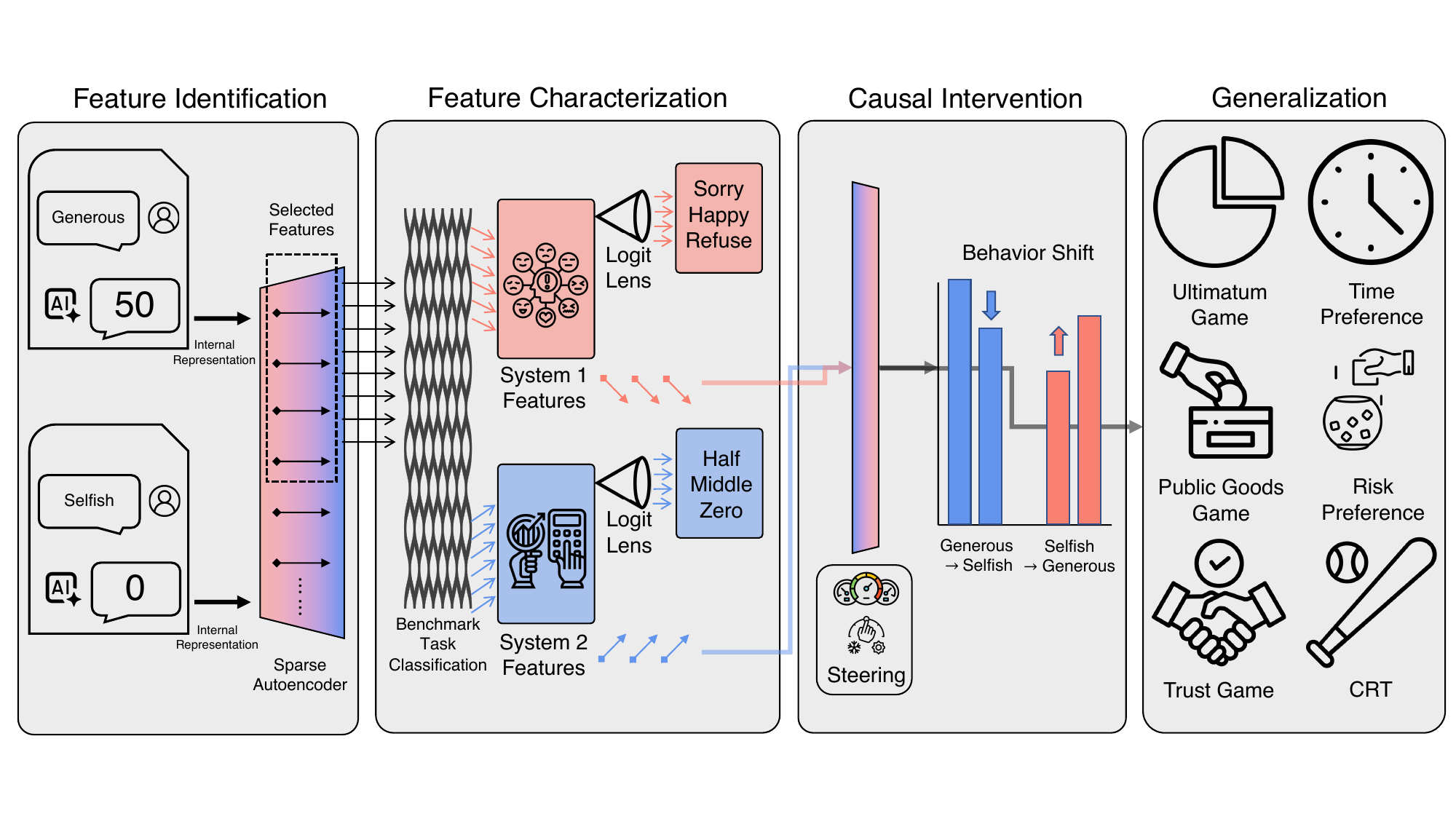}
\caption{\textbf{Overview of the four-stage analysis pipeline.} The pipeline proceeds in four stages. \textbf{(1) Feature identification}: We elicit generous versus selfish behavior in the Dictator Game using minimal-pair prompts (matched prompts that differ only in a single persona word) and compute the resulting difference in internal representations. We then use SAEs to identify features that are consistently associated with this representational difference across prompt variants. \textbf{(2) Feature characterization}: We characterize the identified features by examining whether they are preferentially active on heuristic versus deliberative benchmark tasks (System 1 versus System 2) and by qualitatively inspecting their promoted tokens using \textit{logit lens} analysis. \textbf{(3) Causal intervention}: We intervene on the identified features using two complementary manipulation strategies and test whether changing their activation predictably alters allocations. \textbf{(4) Generalization}: We extend these feature interventions to other decision scenarios to examine whether their effects generalize beyond the Dictator Game, and conduct placebo tests using cognitive reflection tasks. See \hyperref[sec_methods]{Methods} for details.}
\label{fig_framework}
\end{figure*}

Using this setup, we study the model's altruistic output in the Dictator Game, in which a decision-maker is endowed with 100 monetary units and chooses how much to allocate to an anonymous recipient \citep{forsythe1994fairness,camerer2003behavioral}. Because pure self-interest predicts zero transfer, the amount of giving is commonly interpreted as the degree of altruism. Our empirical strategy follows a four-stage pipeline, as illustrated in Figure~\ref{fig_framework}. 

First, we prompt the model to act either generously or selfishly while holding the task otherwise fixed, and compare its internal representations across the two conditions. Second, we use SAEs to identify the features that distinguish these representations and characterize what these features represent. Third, we intervene on the identified features to determine whether they are merely associated with altruistic behavior or causally influence the model’s allocations. Finally, we repeat these interventions in other decision scenarios to examine whether the same features also influence prosocial behavior beyond the Dictator Game.

To the best of our knowledge, this study is among the first mechanism-oriented analyses of prosocial decision-making in LLMs, extending behavioral evaluation by identifying internal features involved in altruistic choices and testing their causal influence on model behavior. These insights may, in turn, inform efforts to make socially relevant LLM behavior more transparent and controllable.

\section{Results}\label{sec_results}

We now present the main results from the four-stage analysis pipeline described above.

\subsection{Identifying features associated with LLMs' altruistic behavior}
We begin by creating a controlled behavioral contrast from which to identify internal features associated with altruism. We do so using matched prompts that assign the model either a generous or a selfish decision-maker persona while holding all other wording fixed. We also include a neutral prompt to benchmark the model's allocations in the absence of a persona instruction.

\begin{figure}[t!]
\centering
\includegraphics[width=0.8\linewidth]{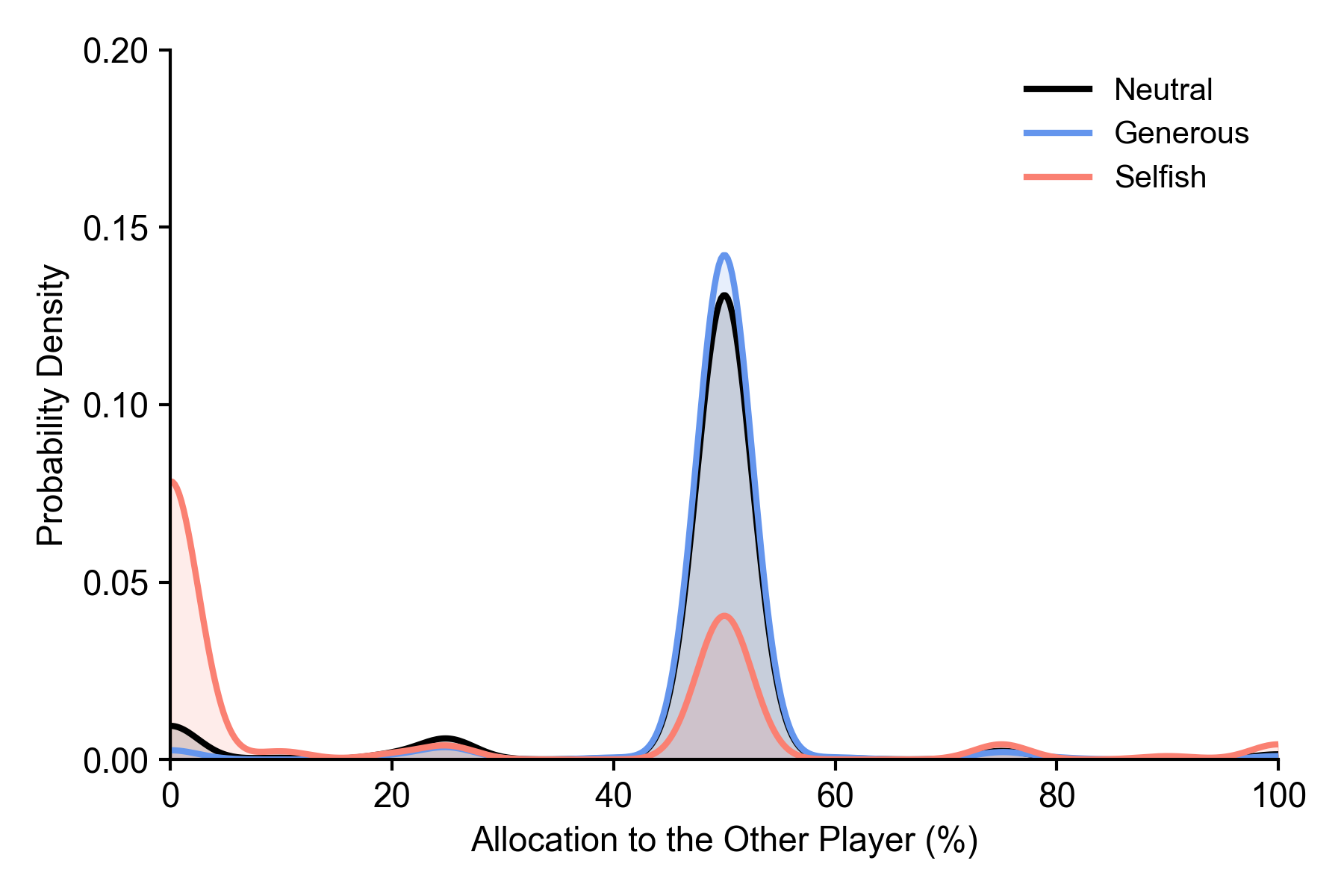}
\caption{\textbf{Altruistic behavior under minimal-pair prompts.} This figure plots the probability density of the proportion of money allocated to the other player. Curves are smoothed using Gaussian kernel density estimation with a fixed bandwidth.} 
\label{fig_baseline}
\end{figure}

Figure~\ref{fig_baseline} reports the distribution of endowment shares allocated to the other player. Under the neutral baseline, the average share allocated to the other player is 46.4\%, close to the levels documented in prior work \citep{mei2024turing,einwiller2025benevolent,kirshner2025prosocial,xie2025using,cook2025what,dodivers2025uncovering}.\footnote{For comparison, prior work finds that human dictators frequently give positive amounts, with an average allocation of approximately 28\% \citep{engel2011dictator}.} Against this benchmark, the generous and selfish persona prompts produce a pronounced contrast in allocation behavior: the average share allocated to the other player is 49.1\% under the generous persona but only 22.1\% under the selfish persona. We observe similar patterns across alternative matched prompt pairs that vary slightly in wording and task parameters (Figure~\ref{fig_minimal_pair_variation}).

Next, we focus on 2,048 candidate SAE features most strongly associated with the contrast between the generous and selfish conditions, representing 0.024\% of all SAE features across the model’s 32 layers.\footnote{The detailed identification procedure is described in \hyperref[sec_methods]{Methods}. The qualitative patterns we report are robust to alternative feature-set sizes.} Because transformer layers process information sequentially, the location of these features indicates where in the model the generous–selfish contrast is most strongly represented. We find that these candidate features are concentrated in the middle layers, with a peak around Layer~20 (Figure~\ref{fig_localization}).

\subsection{Characterizing identified features}
Having identified candidate features associated with altruistic behavior, we next characterize what they represent from two complementary perspectives: the types of tasks in which they are activated and the output tokens they most strongly promote.

We first examine whether the identified features are differentially activated across two broad families of external benchmark tasks: ``System~1'' tasks involving relatively direct associative, affective, and stereotype-related judgments, and ``System~2'' tasks requiring more explicit multi-step reasoning.\footnote{These labels are operational summaries of benchmark activation profiles and do not imply that the model implements human-like dual-process cognition.} We rank the features according to their relative activation across the two task families. Features in the bottom and top tertiles of this ranking are classified as System~1 features (327 features) and System~2 features (328 features), respectively.

\begin{figure}[htbp]
\centering
\includegraphics[width=0.8\linewidth]{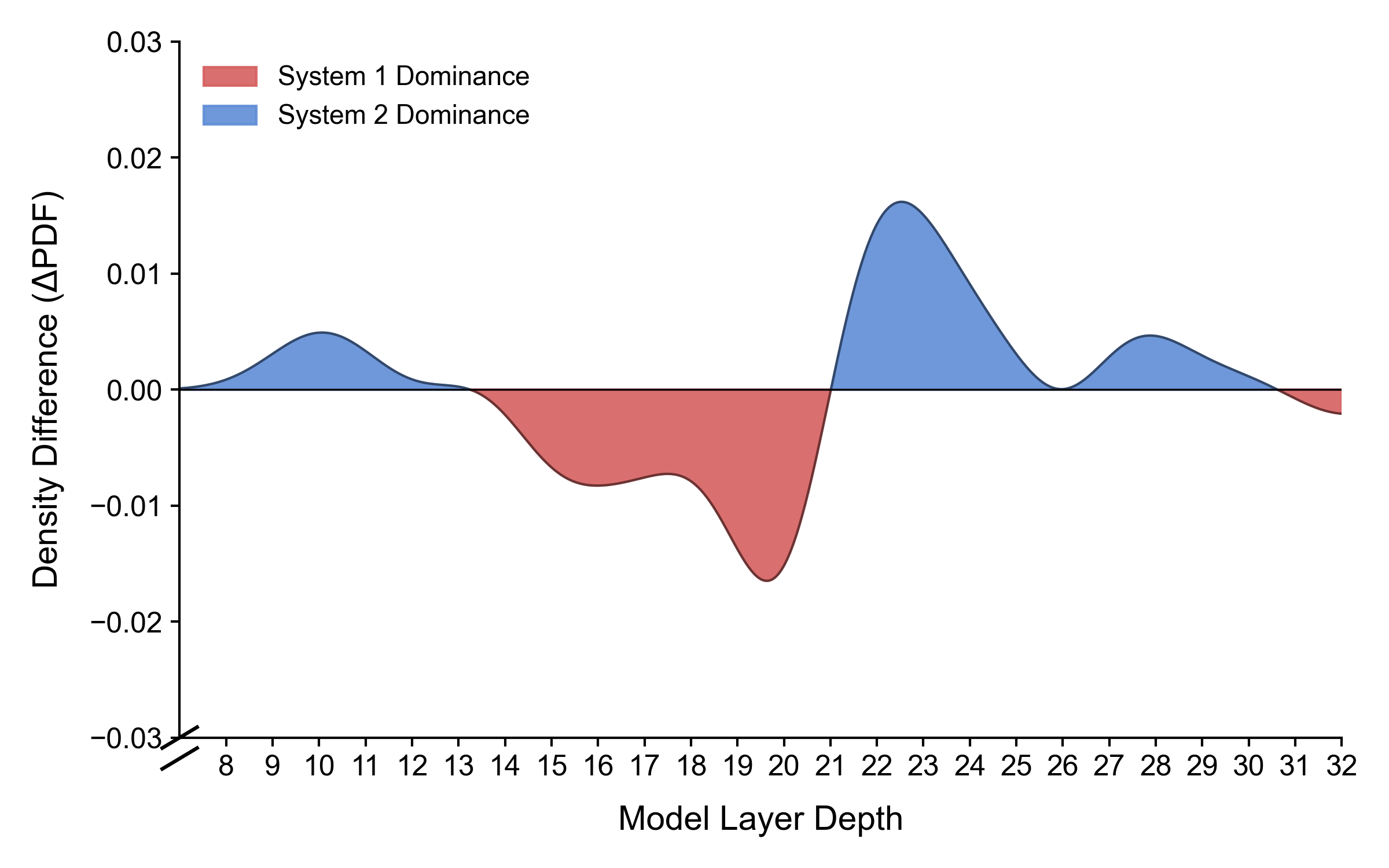}
\caption{\textbf{Layer-wise distribution of System~1 and System~2 features.} The figure shows the difference in density between the two feature groups across model layers. Curves are smoothed using Gaussian kernel density estimation with a fixed bandwidth. The red shaded region (negative values) indicates layers where the density of System~1 features exceeds that of System~2 features. The blue shaded region (positive values) indicates layers where System~2 features are relatively more prevalent.}
\label{fig_sys12_dif_distribution}
\end{figure}

Figure~\ref{fig_sys12_dif_distribution} shows how the two feature groups are distributed across layers. The early layers (before Layer~13) and later layers (after Layer~26) show little difference between the two groups. In the intermediate layers, however, a sequential pattern emerges: System~1 features are more prevalent in Layers~13--20, whereas System~2 features are more prevalent in Layers~21--26.

\begin{figure*}[htbp] 
\centering 
\includegraphics[width=\textwidth]{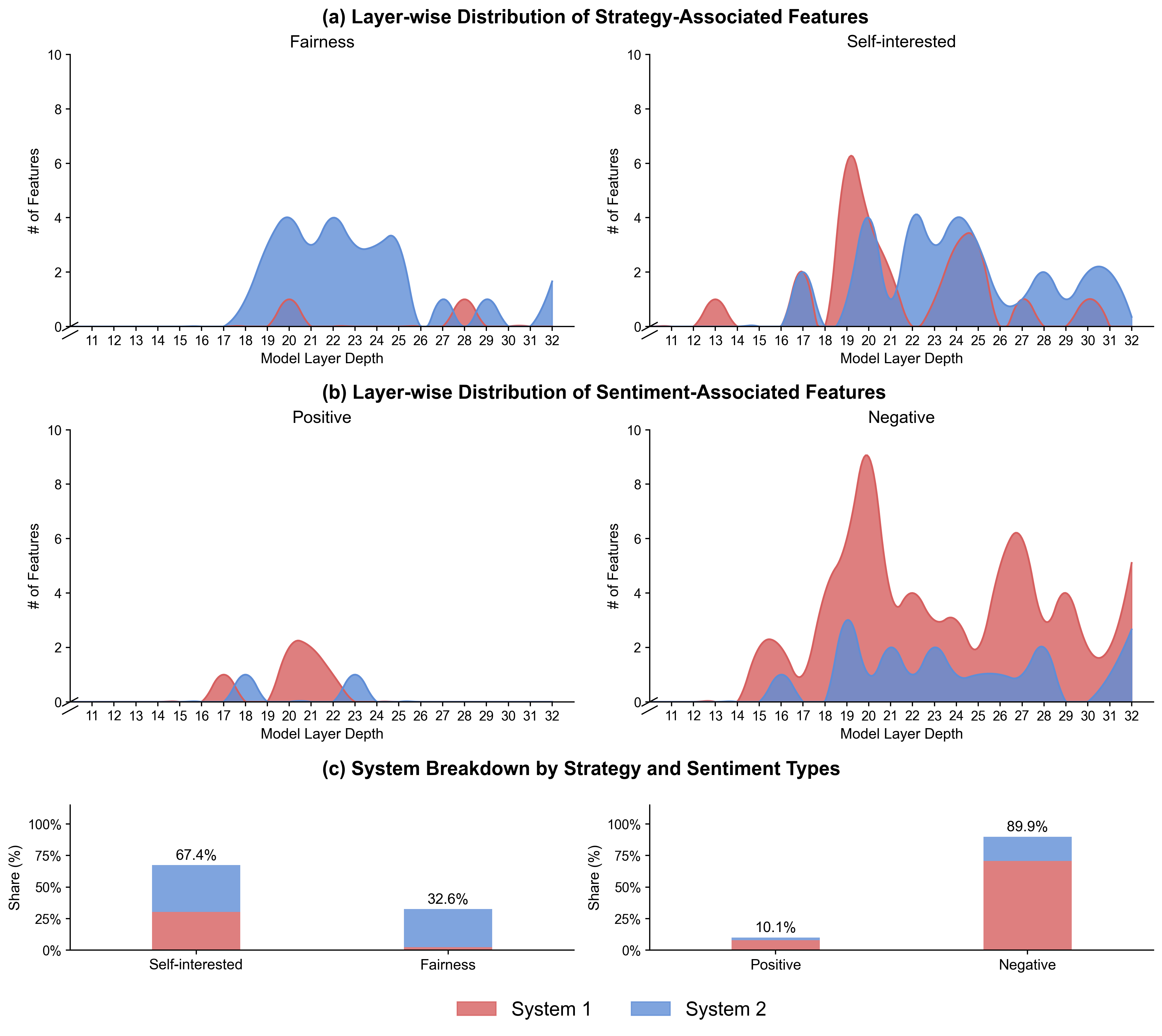} \caption{\textbf{Semantic characterization of identified features.
\label{fig_sys12_logitlens}}
Panels (a) and (b) show the layer-wise distributions of features classified into the Strategy and Sentiment categories, respectively. Strategy features are further classified as Self-interested or Fairness, and Sentiment features as Positive or Negative. Shaded regions distinguish System~1 (red) and System~2 (blue) features; discrete feature counts are smoothed using quadratic B-spline interpolation. Panel (c) summarizes the overall composition of these semantic subcategories. The height of each stacked bar represents the share of features in that subcategory within its broader category (Strategy or Sentiment), while the stacked segments show the shares classified as System~1 and System~2.} 
\end{figure*}

We next examine the semantic content associated with these features using the \textit{logit lens} technique \citep{nostalgebraist2020}, which identifies the output tokens most strongly promoted by each feature. To summarize recurring patterns in these tokens, we use a manually constructed dictionary organized into two broad categories---``Strategy'' and ``Sentiment''---and their respective semantic subcategories (Appendix Section~\ref{appendix_classification_dict}). Strategy features are classified as either Self-interested or Fairness, while Sentiment features are classified as either Positive or Negative.
Figure~\ref{fig_sys12_logitlens} reveals two main patterns, summarized in panel (c). First, among Strategy features, 67.4\% are classified as Self-interested and 32.6\% as Fairness. Self-interested features are distributed across both System~1 and System~2, whereas Fairness features are predominantly classified as System~2. Representative promoted tokens include \textit{zero} and \textit{minimal} for self-interest, and \textit{split} and \textit{equal} for fairness. Second, among Sentiment features, Negative features are much more prevalent than Positive features (89.9\% versus 10.1\%). Representative promoted tokens include \textit{hate} and \textit{sad} for negative sentiment, and \textit{fun} and \textit{exciting} for positive sentiment, as illustrated in Table~\ref{tab_logitlens_examples}. These patterns provide a complementary semantic characterization of the identified features.

\subsection{Causal interventions in the Dictator Game}

We next test whether the identified features are merely associated with allocation outcomes or causally influence them. We first use \textit{activation patching}, an intervention that holds the prompt and model parameters fixed while replacing the activations of selected features with their corresponding values from a contrasting condition. Under the selfish prompt, we replace the activations of the identified features with those from the generous condition; under the generous prompt, we make the reverse replacement. If these features causally influence the model's allocation, patching in generous-condition activations should increase giving under the selfish prompt, whereas the reverse intervention should decrease giving under the generous prompt (see \hyperref[sec_methods]{Methods} for details).

\begin{figure*}[htbp]
\centering
\includegraphics[width=\textwidth]{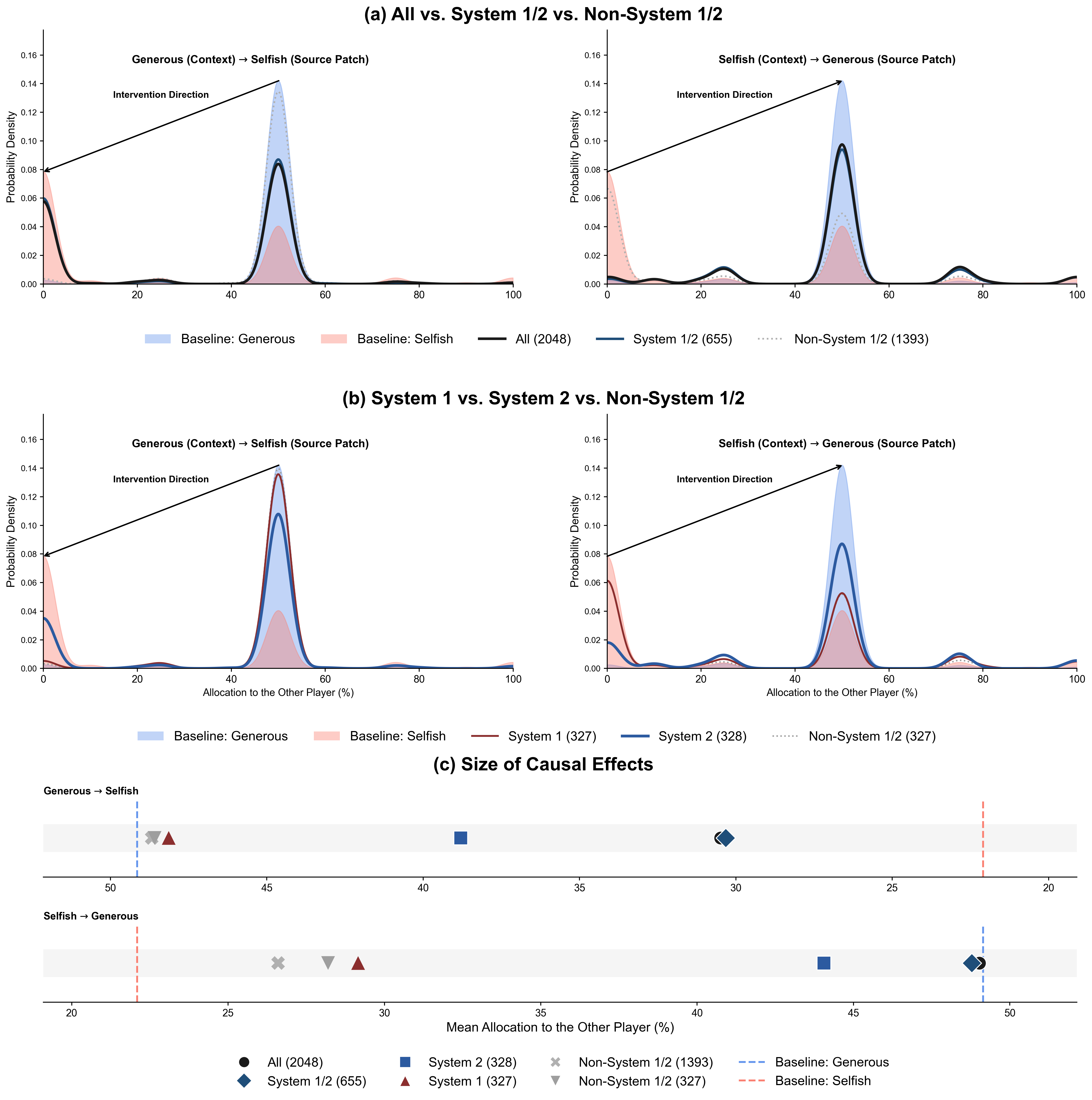}
\caption{\textbf{Causal effects of activation patching in the Dictator Game.}
Panels (a) and (b) show the distributions of allocations produced by activation-patching interventions. Shaded regions indicate the baseline allocation distributions under the generous and selfish prompts, and curves show the resulting distributions after patching different sets of features. The arrows indicate the predicted direction of the behavioral shift. Panel (a) compares interventions on the full set of all 2,048 features, the subset classified as System~1 or System~2, and the remaining unclassified features. Panel (b) separates the effects of System~1 and System~2 features and compares them with a size-matched random sample of unclassified features. Panel (c) summarizes the corresponding mean allocations. Dashed vertical lines mark the pre-intervention baselines under the generous and selfish prompts; larger deviations from these baselines indicate stronger intervention effects.
} 
\label{fig_intervention}
\end{figure*}

Figure~\ref{fig_intervention} shows that activation patching shifts allocations strongly in the predicted directions. Patching the full set of 2{,}048 candidate features moves the allocation distribution toward the contrasting condition, and patching only the 655 features classified as System~1 or System~2 reproduces nearly the full effect (Figure~\ref{fig_intervention}(a)).\footnote{Figure~\ref{fig_topK_robustness} shows that the results are robust to varying the number of candidate features included in the intervention.} In contrast, patching the remaining 1{,}393 unclassified features produces only a small change. Separating the two classified groups reveals that the effect is concentrated among System~2 features (Figure~\ref{fig_intervention}(b--c)). Under the generous prompt, patching System~2 features toward their selfish-condition values reduces mean giving by 10.3 percentage points, compared with 1.0 percentage point for System~1 features and 0.6 percentage points for a size-matched set of randomly selected unclassified features. The reverse intervention under the selfish prompt produces qualitatively similar effects in the opposite direction. These results indicate that System~2 features exert a larger causal influence on allocations in this task. Moreover, interventions on strategy-associated features likewise substantially change allocations, whereas interventions on sentiment-associated features have little effect (Figure~\ref{fig_intervention_semantic_type}).

Activation patching establishes that changing the identified features is sufficient to alter the model's allocations, but it requires activations from a matched contrasting condition. We therefore construct a more general intervention that can be applied without such a comparison. Specifically, we form a steering vector, $\mathbf{f}$, from the generous–selfish activation differences of the System 1 and System 2 features. The vector summarizes the direction from the activation pattern associated with selfish choices toward that associated with generous choices.

We apply this steering vector under a neutral prompt that assigns no persona. As the model generates its response, we shift its internal representation by a scaled version of the vector. The coefficient $\alpha$ determines the direction and magnitude of the intervention: positive values shift activations toward the generous direction, negative values shift them toward the selfish direction, and larger absolute values produce stronger interventions (see \hyperref[sec_methods]{Methods} for details). Figure~\ref{fig_neutral_steering} shows a monotonic behavioral response to $\alpha$: positive values shift allocations toward an equal split, whereas negative values shift them toward zero giving.

\subsection{Generalization and placebo tests}

We next ask whether the steering vector derived from the Dictator Game generalizes to other social decision settings. We apply the same steering vector, $\mathbf{f}$, in four additional settings: the Dictator Game with modified transfer multipliers, the Ultimatum Game in both proposer and responder roles, the Public Goods Game, and the Trust Game (see \hyperref[sec_methods]{Methods} for details).

\begin{figure*}[htbp]
\centering
\includegraphics[width=\textwidth]{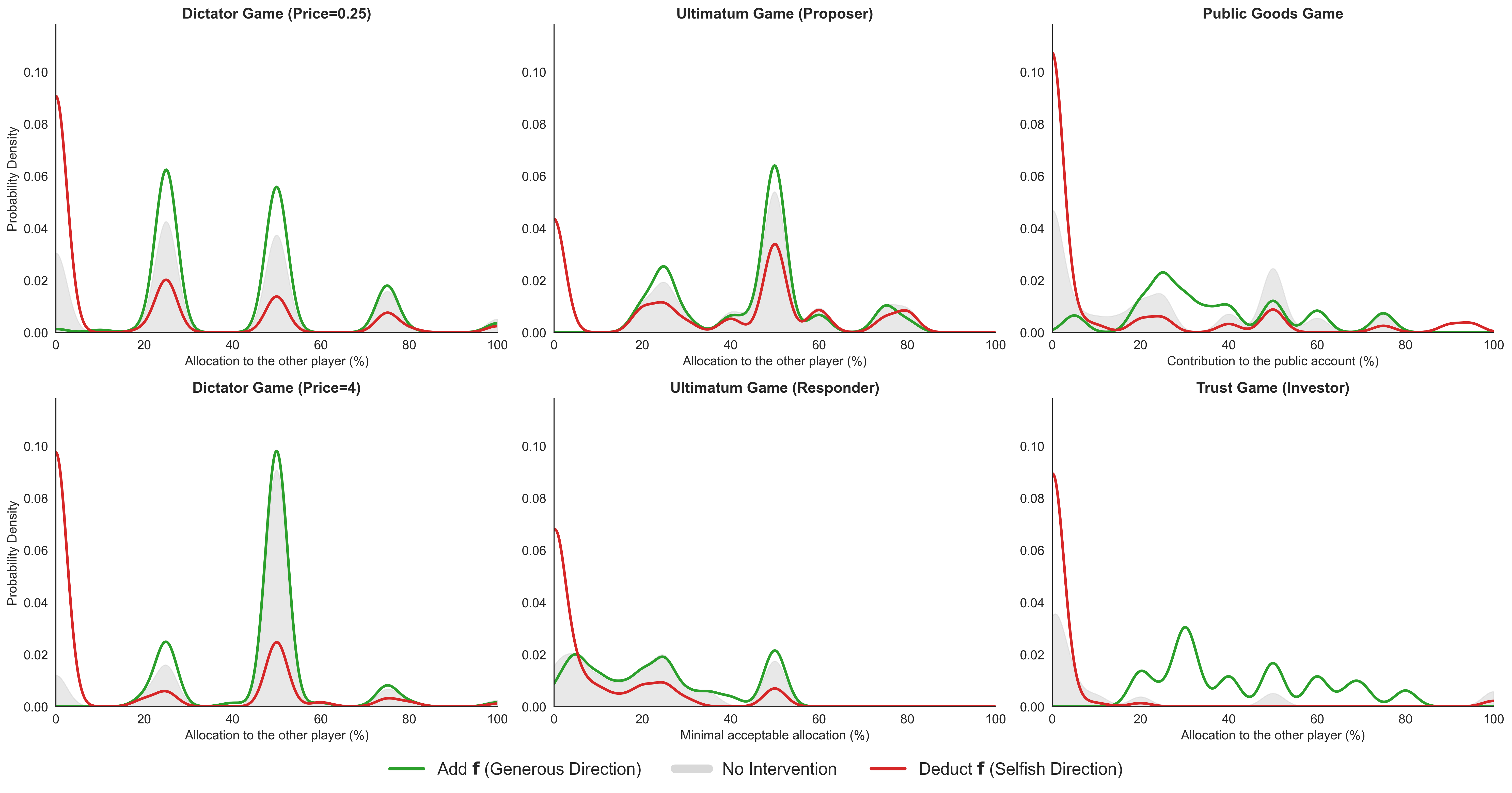}
\caption{\textbf{Steering experiments across four other economic settings.} 
The steering vector $\mathbf{f}$, originally derived from the Dictator Game, is applied to four other contexts: the Dictator Game with transfer multipliers (0.25 and 4), the Ultimatum Game (proposer and responder), the Public Goods Game, and the Trust Game (investor role).
Shaded gray regions represent the distribution of baseline responses before intervention. 
Solid green and red lines indicate the response distributions after applying the steering vector in the positive (generous-aligned, $\alpha=1$) and negative (selfish-aligned, $\alpha=-1$) directions, respectively. 
Curves are smoothed using Gaussian kernel density estimation with a fixed bandwidth.}
\label{fig_external_validity}
\end{figure*}

Figure~\ref{fig_external_validity} shows broadly consistent effects across these settings. Positive steering ($\alpha=1$) shifts behavior in the generous or fairness-oriented direction, whereas negative steering ($\alpha=-1$) shifts behavior in the opposite direction. In the Dictator, Public Goods, and Trust Games, positive steering increases allocations, contributions, and investments, respectively, whereas negative steering decreases them. A similar directional pattern emerges in the Ultimatum Game for both proposer offers and responders’ minimum acceptable offers. The magnitude of these effects can be substantial. In the Trust Game, for instance, when the model acts as the investor, average investment increases from 16.4\% of the endowment without intervention to 43.2\% under positive steering, but falls to 2.8\% under negative steering. Figure~\ref{fig_time_risk} reports analogous steering experiments in risk- and time-preference tasks.

Finally, we conduct placebo tests to examine whether the steering vector induces a generic shift in numerical responses rather than specifically affecting socially relevant behavior. We apply the same steering vector, $\mathbf{f}$, to three Cognitive Reflection Test (CRT) questions \citep{frederick2005cognitive,hagendorff2023human}, which also require numerical answers but do not involve social preferences. Across all three questions, steering produces negligible changes (Figure~\ref{fig_placebo}), suggesting that the intervention does not simply alter numerical outputs.

\section{Discussion}\label{sec_discussion}

This study moves beyond behavioral descriptions of altruism in LLMs by showing that internal features associated with altruistic choices can be identified and causally manipulated. Intervening on these features predictably changes allocations, and feature-based interventions produce consistent directional effects across several social decision settings.

Prior work has shown that LLMs can exhibit human-like economic preferences and biases \citep{chen2023emergence,mei2024turing,akata2025playing,johnson2025testing,xie2025using,goli2024frontiers,salecha2024large,macmillan2024ir,bini2025behavioral}. In the domain of social preferences, models often display prosocial behavior, but their choices can vary substantially with prompts and task structure \citep{einwiller2025benevolent,kirshner2025prosocial}. Most of this literature characterizes models from their outputs, leaving open which internal processes contribute to these behavioral patterns. More broadly, a growing mechanistic interpretability literature examines how internal representations contribute to LLM outputs \citep{gantla2025exploring,lindsey2025biology,chen2025financial}. Within the domain of social preferences, recent work has begun to manipulate internal representations to alter prosocial choices \citep{ma2025computational}. Our findings advance this line of work by providing causal evidence that specific internal representations contribute to prosocial decision-making and by introducing a general framework that uses behavioral contrasts to identify relevant internal features and interventions to test their causal influence on model behavior.

Beyond their causal relevance, the identified internal features also exhibit substantial heterogeneity. Features associated with heuristic versus deliberative benchmark tasks differ both in where they appear in the model and in how strongly their manipulation affects behavior. The \textit{logit lens} analysis further reveals systematic semantic differences. Fairness-related features are concentrated among those associated with deliberative benchmarks, whereas self-interested features span both groups. This pattern may suggest that fairness is less closely associated with default or automatic responses and more closely associated with representations active during deliberative processing. Such an interpretation is consistent with cognitive control accounts of human altruism, which propose that prosocial behavior can require executive control to override self-interested impulses \citep{steinbeis2012impulse}. We also find that features associated with negative sentiment are considerably more common than those associated with positive sentiment, which may suggest that these representations are more closely related to avoiding negative affective states, such as guilt, than to pursuing positive moral satisfaction.

Finally, our generalization tests show that a steering vector derived from the non-strategic Dictator Game transfers to other social decision environments. Unlike the Dictator Game, several of these settings introduce strategic incentives and, in some cases, expectations about another player's response. The similar effects across games are therefore consistent with the model reusing some of the shared internal representations across different decision contexts \citep{hong2024implies,du2025human,vafa2025what,li2025geometry}.

Several limitations warrant acknowledgment. Although we identify features that causally influence prosocial choices, we do not fully characterize how these features interact with one another or with other components of the model \citep{wang2023interpretability,hong2024implies}. They may therefore capture only part of the processes that generate prosocial choices. In addition, our characterization of these features relies on their activation patterns in benchmark tasks and their associations with output tokens. The resulting System~1/System~2 and semantic labels are descriptive and do not establish that the model implements corresponding human-like cognitive or neural processes. Addressing these limitations will require mapping how the identified features work together and more closely connecting analyses of LLM internal representations with research on human cognition. Such work would provide a stronger foundation for explaining, evaluating, and steering LLM behavioral outputs.

\newpage
\section{Methods}\label{sec_methods}

Throughout the study, we use Llama-3.1-8B-Instruct, an open-weight LLM, to generate choices and OpenSAE, an open-source SAE toolkit, to analyze and intervene on the model's internal features. Our analysis follows the four-stage pipeline described in Figure~\ref{fig_framework}, linking prosocial behavior to internal features and testing their causal influence on model behavior.

\paragraph*{The Dictator Game} 
We measure altruism using the Dictator Game, which is widely used in human experiments \citep{camerer2003behavioral} and has recently been adapted to LLMs \citep{mei2024turing}. The model is endowed with a budget (e.g., \$100) and asked to choose how much to transfer to an anonymous recipient (prompt in Appendix Section~\ref{appendix_prompts_dictator}). Rather than relying on a single decoded response, we measure the model's allocation from its next-token probability distribution. We retain the ten most likely next tokens, identify those corresponding to valid transfers between 0 and the endowment, normalize their probabilities, and compute the resulting probability-weighted mean allocation.

\paragraph*{Minimal-pair prompts}
To create a clean behavioral contrast while holding the economic environment fixed, we use a minimal-pair prompt design (Appendix Section~\ref{appendix_prompts_baseline}). Originating in linguistics \citep{chomsky1968sound} and later adapted for neural probing \citep{warstadt2020blimp,ettinger2020bert}, this approach compares inputs that differ in only one targeted element. In our setting, the two prompts differ only in a single persona word:
\begin{itemize}
    \item Generous condition: ``You are a \textbf{\textit{generous}} decision maker ..."
    \item Selfish condition: ``You are a \textbf{\textit{selfish}} decision maker ...”
\end{itemize}
Because all other wording is held fixed, this design minimizes textual differences between conditions and allows us to link differences in internal activity to the persona manipulation.

\paragraph*{Feature identification} 

Denote the above minimal pair by $p_0 = (x_{\text{gen}}^{0}, x_{\text{self}}^{0})$. Let $\mathbf{h}^{(l)}(x)$ denote the model's internal representation at the final token of input prompt $x$ in layer $l$, where $l=1,\dots,L$ and $L$ is the number of layers. We calculate the generous--selfish difference at layer $l$ as follows:
$$
\boldsymbol{\delta}_{p_0}^{(l)}
=
\mathbf{h}^{(l)}(x^0_{\text{gen}})
-
\mathbf{h}^{(l)}(x^0_{\text{self}}).
$$

We use SAEs to represent $\mathbf{h}^{(l)}(x)$ as an approximately sparse linear combination of learned features \citep{bricken2023monosemanticity}, which are usually easier to interpret.
Let $\mathcal{I}^{(l)}$ denote the set of SAE features in layer $l$, with feature indices distinct across layers. For feature $i \in \mathcal{I}^{(l)}$, $\mathbf{d}_i^{(l)}$ denotes its unit-length direction, and $f_i(\mathbf{h}^{(l)}(x)) \geq 0$ denotes its activation magnitude when processing prompt $x$. 
To identify the features most strongly aligned with the generous--selfish representational contrast, we define an attribution score:
$$
A_{i,p_0}^{(l)}
=
\bigl(\boldsymbol{\delta}_{p_0}^{(l)}\bigr)^{\top}
\mathbf{d}_i^{(l)}.
$$
The score measures the magnitude and direction of the representational difference along each feature direction.

\begin{figure*}[htbp]
\centering
\includegraphics[width=\textwidth]{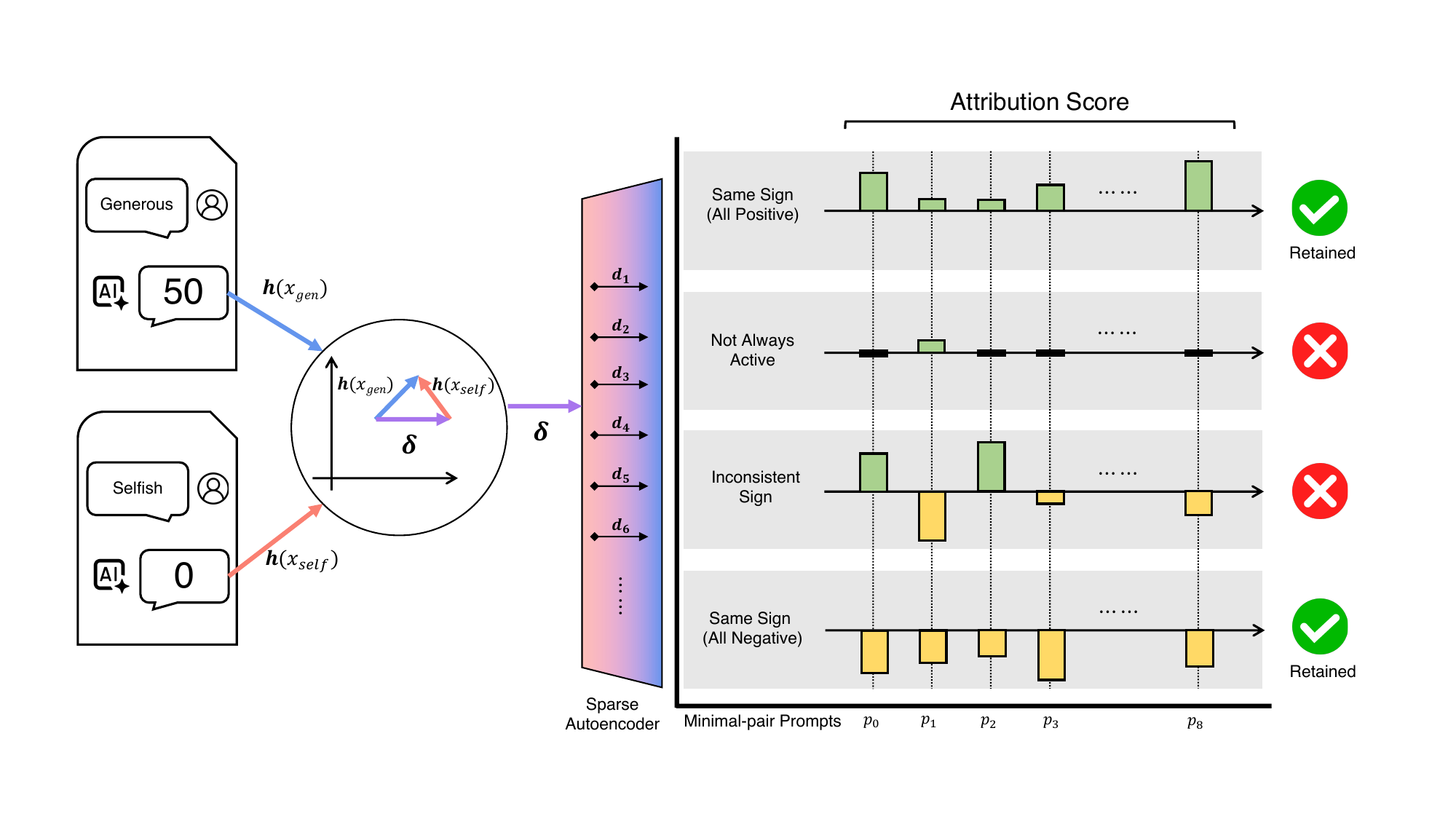}
\caption{\textbf{Overview of the feature identification process.} We identify candidate features associated with altruistic behavior. First, internal representations are extracted from the model under minimal-pair prompts (generous versus selfish) in the Dictator Game. At each layer, the generous--selfish difference vector, $\boldsymbol{\delta}=\mathbf{h}(x_{\text{gen}})-\mathbf{h}(x_{\text{self}})$, is computed and projected onto each SAE feature direction to obtain an attribution score. Features are then screened using the baseline pair and eight prompt variations ($p_0,\dots,p_8$). Within each layer and prompt pair, we shortlist up to 1,000 active features with the largest nonzero absolute attribution scores. A feature passes the consistency screen if it appears in the shortlist for every prompt pair and its attribution retains the same sign across all variations. Features that do not appear in the shortlist for one or more prompt pairs or have inconsistent attribution signs are discarded. The retained features are subsequently ranked by their average absolute attribution to form the candidate feature set.}
\label{fig_identification}
\end{figure*}

To reduce sensitivity to a single wording choice, we also use a validation set of minimal pairs $\mathcal{P} = \{p_1, \dots, p_K\}$ (prompts in Section~\ref{appendix_prompts_variation}) with $K=8$ variations (e.g., different endowments such as \$20, \$70, and \$128; alternative task wording such as allocating versus receiving; and adjective synonyms such as ``prosocial'' versus ``egoistic''). 
For each validation pair $p_k=(x^k_{\text{gen}},x^k_{\text{self}})$, we calculate $\boldsymbol{\delta}_{p_k}^{(l)}$ and $A_{i,p_k}^{(l)}$ in the same way.
For $k=0,\dots,K$, a feature is active in pair $p_k$ when
$|f_i(\mathbf{h}^{(l)}(x^k_{\text{gen}}))| + |f_i(\mathbf{h}^{(l)}(x^k_{\text{self}}))| > 0$.
Within each layer and prompt pair, we shortlist up to 1,000 active features with the largest nonzero absolute attribution scores. 
We retain only features that appear in the shortlist for every pair in $\{p_0\}\cup\mathcal{P}$ and whose attribution has a consistent sign, i.e.,
$\operatorname{sign}(A_{i,p_0}^{(l)}) = \operatorname{sign}(A_{i,p_k}^{(l)})$
for all $p_k \in \mathcal{P}$.
This screening procedure is designed to retain features whose association with the generous--selfish contrast is stable across prompt variations. Finally, we rank the retained features by their average absolute attribution across $\{p_0\} \cup \mathcal{P}$ and select the top-ranked ones to form the candidate set $\widetilde{\mathcal{I}}$. We denote the selected features in layer $l$ by $\widetilde{\mathcal{I}}^{(l)}$. Figure~\ref{fig_identification} illustrates the feature identification process.

\subsection*{Classifying System 1 and System 2 features}\label{sec_classification}

To characterize the candidate features, we examine their activation across six standard benchmark tasks, grouped into two broad families. We refer to tasks involving relatively direct associative, affective, or stereotype-related judgments as ``System~1'' tasks, and tasks requiring more explicit multi-step reasoning as ``System~2'' tasks. The System~1 set ($\mathcal{T}_{S1}$) includes TruthfulQA \citep{lin2022truthfulqa}, Crowdsourced Stereotype Pairs (CrowS-Pairs) \citep{nangia2020crows}, and Stanford Sentiment Treebank (SST-2) \citep{socher2013recursive}. The System~2 set ($\mathcal{T}_{S2}$) includes Grade School Math 8K (GSM8K) \citep{cobbe2021gsm8k}, Big-Bench Hard (BBH) \citep{suzgun2023challenging}, and StrategyQA \citep{geva2021did}.

We sample $N=30$ questions from each benchmark (sample questions in Appendix Section~\ref{appendix_benchmark}), yielding 90 questions in each task family. For each candidate feature $i\in\widetilde{\mathcal{I}}$, let $C_{i,t}$ denote the number of questions in benchmark task $t$ for which the feature is active. We aggregate these counts within each task family as
$S_{i,S1} = \sum_{t\in\mathcal{T}_{S1}} C_{i,t}$ and $S_{i,S2} = \sum_{t\in\mathcal{T}_{S2}} C_{i,t}$,
and for features with $S_{i, S2} + S_{i, S1} > 0$, measure the feature's relative activation across the two task families as
$D_i = \frac{S_{i, S2} - S_{i, S1}}{S_{i, S2} + S_{i, S1}},$ where $D_i\in[-1,1]$. Positive values indicate relatively greater activation on System~2 tasks, whereas negative values indicate relatively greater activation on System~1 tasks; $D_i=1$ ($-1$) indicates activation only on System~2 (System~1) tasks. We retain only features with sufficient activity in their more active task family. Specifically, a feature must be active in at least 3 of the 30 sampled questions (10\%) in at least one benchmark from that family. We then rank the retained features by $D_i$ and classify those in the bottom, middle, and top tertiles as ``System~1,'' ``Ambiguous,'' and ``System~2'' features, respectively.

\subsection*{Establishing causality between features and altruistic behavior} \label{sec_intervention}
We test whether the identified features play a causal role by intervening on them in two ways. We exclude features in the first and final layers from these tests because these boundary layers have distinct functional roles: early layers are closely tied to input processing, whereas the final layer is directly coupled to the model's output logits. We therefore focus on intermediate layers, where internal representations are more fully developed and can be manipulated before the model produces its final output \citep{meng2022locating,geva2022transformer,zou2023representation}.\footnote{For consistency with the feature-identification results, we report feature-set sizes before excluding the first and final layers; the interventions themselves are applied only to features in the intermediate layers.}

\paragraph*{Activation patching with minimal-pair prompts} 
Our first causal test uses \textit{activation patching} \citep{meng2022locating,wang2023interpretability,wu2024mitigating}, which replaces selected feature activations in one condition with the corresponding activations from another condition while holding the prompt and model parameters fixed. We run the Dictator Game under both prompts in the minimal pair $p_0$. Under the selfish prompt, we replace the activations of the identified features with those recorded at the same layer and token position under the generous prompt. We then reverse the procedure by replacing the corresponding activations under the generous prompt with those from the selfish condition. If patching generous-condition activations into the selfish condition increases giving, while the reverse intervention decreases it, this would show that selectively changing these features is sufficient to systematically alter the model's choices and provide causal evidence for their role in altruistic behavior.

\paragraph*{Bidirectional steering with neutral prompts}

Activation patching uses feature values observed under the two persona prompts. We next use the generous–selfish activation differences of the candidate features classified as System 1 or System 2 to construct an intervention that can be applied when the prompt assigns no persona (Appendix Section~\ref{appendix_prompts_dictator}). Specifically, for each such feature $i$, we compute its activation difference between the minimal-pair prompts $p_0$,
$$\Delta f_i = f_i(\mathbf{h}^{(l)}(x^0_{\text{gen}})) - f_i(\mathbf{h}^{(l)}(x^0_{\text{self}})).$$
We combine these differences along the corresponding feature directions $\mathbf{d}_i^{(l)}$ to construct a layer-specific steering vector:
$$
\begin{aligned}
\mathbf{f}^{(l)}
&=
\sum_{i}
\Delta f_i\,\mathbf{d}_i^{(l)},\\
\mathbf{h}'^{(l)}
&=
\mathbf{h}^{(l)}+\alpha\mathbf{f}^{(l)},
\end{aligned}
$$
where the sum runs over the System 1 and System 2 features in layer \(l\).

The parameter $\alpha$ controls the direction and magnitude of the intervention. When $\alpha=0$, the model is unsteered. Positive values shift the model state in the direction associated with moving from the selfish to the generous condition, whereas negative values shift it in the opposite direction. We test whether increasing $\alpha$ shifts allocations toward larger transfers and decreasing $\alpha$ shifts them toward smaller transfers relative to the unsteered neutral condition. The generalization analyses below apply the same steering vectors.

\paragraph*{Generalization to other decision settings} 
We test whether the steering vector derived from the Dictator Game also shifts behavior in other economic decision settings. In every test, we use the same feature set and layer-specific steering vectors $\mathbf{f}^{(l)}$. We further consider four economic games: (1) Dictator Games with transfer multipliers of 0.25 or 4 (with similar results for alternative scalings of 0.1, 0.5, 2, and 10), (2) the Public Goods Game, (3) the Ultimatum Game in both proposer and responder roles, and (4) the Trust Game in the investor role. Motivated by evidence that social preferences may correlate with risk and time preferences \citep{burks2009cognitive,dohmen2011individual,chen2025general}, we also examine standard budgetary decision tasks for risk and time preferences \citep{chen2023emergence,choi2007consistency}. The prompts are reported in Appendix Section~\ref{appendix_prompts_external_validity}.

\paragraph*{Placebo tests} 
To test whether the steering intervention simply changes numerical responses, we apply the same steering vector to three Cognitive Reflection Test (CRT) questions \citep{frederick2005cognitive,hagendorff2023human} (prompts in Section~\ref{appendix_prompts_placebo}). CRT questions are short numerical problems designed to elicit an intuitive but incorrect answer that can be corrected through further reflection. Like the Dictator Game, they require integer responses, but they do not involve resource sharing or social-preference tradeoffs. If the steering vector does not simply induce a generic shift in numerical responses, it should have little effect on the distribution of CRT answers.
\newpage

\clearpage
\singlespacing
\bibliographystyle{aea}
\bibliography{ref.bib}

@article{hsu2015,
  title={Dopamine modulates egalitarian behavior in humans},
  author={Sáez, Ignacio and Lusha, Zhu and Eric, Set and Andrew, Kayser and and Ming, Hsu},
  journal={Current Biology},
  volume={25},
  number={7},
  pages={912-919},
  year={2015},
  publisher={Oxford University Press}
}

@article{Awad2018,
  title={The moral machine experiment},
  author={Awad, Edmond and Sohan, Dsouza and Richard, Kim and Jonathan, Schulz and Joseph, Henrich and Azim, Shariff and Jean-François, Bonnefon and Iyad and Rahwan},
  journal={Nature},
  volume={563},
  number={7729},
  pages={59-64},
  year={2018},
  publisher={Oxford University Press}
}

@article{choi2007consistency,
  title     = {Consistency and heterogeneity of individual behavior under uncertainty},
  author    = {Choi, Syngjoo and Fisman, Raymond and Gale, Douglas and Kariv, Shachar},
  journal   = {American Economic Review},
  volume    = {97},
  number    = {5},
  pages     = {1921--1938},
  year      = {2007}
}

@article{macmillan2024ir,
  title     = {({I}r)rationality and cognitive biases in large language models},
  author    = {Macmillan-Scott, Olivia and Musolesi, Mirco},
  journal   = {Royal Society Open Science},
  volume    = {11},
  number    = {6},
  pages     = {240255},
  year      = {2024},
  publisher = {The Royal Society}
}

@article{frederick2005cognitive,
  title     = {Cognitive reflection and decision making},
  author    = {Frederick, Shane},
  journal   = {Journal of Economic Perspectives},
  volume    = {19},
  number    = {4},
  pages     = {25--42},
  year      = {2005},
  publisher = {American Economic Association}
}

@article{goli2024frontiers,
  title     = {Frontiers: Can large language models capture human preferences?},
  author    = {Goli, Ali and Singh, Amandeep},
  journal   = {Marketing Science},
  volume    = {43},
  number    = {4},
  pages     = {709--722},
  year      = {2024},
  publisher = {INFORMS}
}

@article{chen2023emergence,
  title     = {The emergence of economic rationality of {GPT}},
  author    = {Chen, Yiting and Liu, Tracy Xiao and Shan, You and Zhong, Songfa},
  journal   = {Proceedings of the National Academy of Sciences},
  volume    = {120},
  number    = {51},
  pages     = {e2316205120},
  year      = {2023},
  publisher = {National Academy of Sciences}
}

@article{mei2024turing,
  title     = {A {T}uring test of whether {AI} chatbots are behaviorally similar to humans},
  author    = {Mei, Qiaozhu and Xie, Yutong and Yuan, Walter and Jackson, Matthew O.},
  journal   = {Proceedings of the National Academy of Sciences},
  volume    = {121},
  number    = {9},
  pages     = {e2313925121},
  year      = {2024},
  publisher = {National Academy of Sciences}
}

@article{hagendorff2023human,
  title     = {Human-like intuitive behavior and reasoning biases emerged in large language models but disappeared in {ChatGPT}},
  author    = {Hagendorff, Thilo and Fabi, Sarah and Kosinski, Michal},
  journal   = {Nature Computational Science},
  volume    = {3},
  number    = {10},
  pages     = {833--838},
  year      = {2023},
  publisher = {Nature Publishing Group US New York}
}

@article{brynjolfsson2025generative,
  title     = {Generative {AI} at work},
  author    = {Brynjolfsson, Erik and Li, Danielle and Raymond, Lindsey},
  journal   = {Quarterly Journal of Economics},
    volume = {140},
    number = {2},
    pages = {889-942},
    year = {2025},
    month = {05},
  publisher = {Oxford University Press}
}

@article{dohmen2011individual,
  title={Individual risk attitudes: Measurement, determinants, and behavioral consequences},
  author={Dohmen, Thomas and Falk, Armin and Huffman, David and Sunde, Uwe and Schupp, J{\"u}rgen and Wagner, Gert G},
  journal={Journal of the European Economic Association},
  volume={9},
  number={3},
  pages={522--550},
  year={2011},
  publisher={Oxford University Press}
}

@article{einwiller2025benevolent,
  title={Benevolent Dictators? {On} {LLM} Agent Behavior in Dictator Games},
  author={Einwiller, Andreas and Dastidar, Kanishka Ghosh and Romazanov, Artur and Hautli-Janisz, Annette and Granitzer, Michael and Lemmerich, Florian},
  journal={arXiv preprint arXiv:2511.08721},
  year={2025}
}

@article{kirshner2025prosocial,
  title={Prosocial When Simple and Cold-Hearted When Complex: How Task Difficulty Shapes {LLM} Behavior},
  author={Kirshner, Samuel N. and Pan, Yiwen and Wu, Jason Xianghua},
  journal={Decision Analysis},
  year={2025},
  doi={10.1287/deca.2025.0396},
  publisher={INFORMS}
}

@article{cobbe2021gsm8k,
  title     = {Training verifiers to solve math word problems},
  author    = {Cobbe, Karl and Kosaraju, Vineet and Bavarian, Mohammad and Chen, Mark and Jun, Heewoo and Kaiser, Lukasz and Plappert, Matthias and Tworek, Jerry and Hilton, Jacob and Nakano, Reiichiro and Hesse, Christopher and Schulman, John},
  journal   = {arXiv preprint arXiv:2110.14168},
  year      = {2021}
}

@article{bricken2023monosemanticity,
title={Towards Monosemanticity: Decomposing Language Models With Dictionary Learning},
author={Bricken, Trenton and Templeton, Adly and Batson, Joshua and Chen, Brian and Jermyn, Adam and Conerly, Tom and Turner, Nick and Anil, Cem and Denison, Carson and Askell, Amanda and Lasenby, Robert and Wu, Yifan and Kravec, Shauna and Schiefer, Nicholas and Maxwell, Tim and Joseph, Nicholas and Hatfield-Dodds, Zac and Tamkin, Alex and Nguyen, Karina and McLean, Brayden and Burke, Josiah E and Hume, Tristan and Carter, Shan and Henighan, Tom and Olah, Christopher},
year={2023},
journal={Transformer Circuits Thread},
url={https://transformer-circuits.pub/2023/monosemantic-features/index.html}
}

@article{templeton2024scaling,
title={Scaling Monosemanticity: Extracting Interpretable Features from {Claude 3 Sonnet}},
author={Templeton, Adly and Conerly, Tom and Marcus, Jonathan and Lindsey, Jack and Bricken, Trenton and Chen, Brian and Pearce, Adam and Citro, Craig and Ameisen, Emmanuel and Jones, Andy and Cunningham, Hoagy and Turner, Nicholas L and McDougall, Callum and MacDiarmid, Monte and Freeman, C. Daniel and Sumers, Theodore R. and Rees, Edward and Batson, Joshua and Jermyn, Adam and Carter, Shan and Olah, Chris and Henighan, Tom},
year={2024},
journal={Transformer Circuits Thread},
url={https://transformer-circuits.pub/2024/scaling-monosemanticity/index.html}
}

@article{lieberum2024gemma,
  title={{Gemma Scope}: Open sparse autoencoders everywhere all at once on {G}emma 2},
  author={Lieberum, Tom and Rajamanoharan, Senthooran and Conmy, Arthur and Smith, Lewis and Sonnerat, Nicolas and Varma, Vikrant and Kram{\'a}r, J{\'a}nos and Dragan, Anca and Shah, Rohin and Nanda, Neel},
  journal={arXiv preprint arXiv:2408.05147},
  year={2024}
}

@article{lindsey2025biology,
  author={Lindsey, Jack and Gurnee, Wes and Ameisen, Emmanuel and Chen, Brian and Pearce, Adam and Turner, Nicholas L. and Citro, Craig and Abrahams, David and Carter, Shan and Hosmer, Basil and Marcus, Jonathan and Sklar, Michael and Templeton, Adly and Bricken, Trenton and McDougall, Callum and Cunningham, Hoagy and Henighan, Thomas and Jermyn, Adam and Jones, Andy and Persic, Andrew and Qi, Zhenyi and Thompson, T. Ben and Zimmerman, Sam and Rivoire, Kelley and Conerly, Thomas and Olah, Chris and Batson, Joshua},
  title={On the Biology of a Large Language Model},
  journal={Transformer Circuits Thread},
  year={2025},
  url={https://transformer-circuits.pub/2025/attribution-graphs/biology.html}
}

@article{li2025geometry,
  title={The geometry of concepts: Sparse autoencoder feature structure},
  author={Li, Yuxiao and Michaud, Eric J and Baek, David D and Engels, Joshua and Sun, Xiaoqing and Tegmark, Max},
  journal={Entropy},
  volume={27},
  number={4},
  pages={344},
  year={2025},
  publisher={MDPI}
}

@misc{opensae,
  title={{OpenSAE}: Open-sourced Sparse Auto-Encoder towards Interpreting Large Language Models},
  author={THU-KEG},
  year={2025},
  url={https://github.com/THU-KEG/OpenSAE}
}

@article{he2024llama,
  title={{Llama Scope}: Extracting millions of features from {Llama-3.1-8B} with sparse autoencoders},
  author={He, Zhengfu and Shu, Wentao and Ge, Xuyang and Chen, Lingjie and Wang, Junxuan and Zhou, Yunhua and Liu, Frances and Guo, Qipeng and Huang, Xuanjing and Wu, Zuxuan and others},
  journal={arXiv preprint arXiv:2410.20526},
  year={2024}
}

@article{du2025human,
  title={Human-like object concept representations emerge naturally in multimodal large language models},
  author={Du, Changde and Fu, Kaicheng and Wen, Bincheng and Sun, Yi and Peng, Jie and Wei, Wei and Gao, Ying and Wang, Shengpei and Zhang, Chuncheng and Li, Jinpeng and others},
  journal={Nature Machine Intelligence},
  volume={7},
  pages={860--875},
  year={2025},
  publisher={Nature Publishing Group UK London}
}

@misc{vafa2025what,
title={What Has a Foundation Model Found? {I}nductive Bias Reveals World Models},
author={Keyon Vafa and Peter G. Chang and Ashesh Rambachan and Sendhil Mullainathan},
howpublished={Proceedings of the 42nd International Conference on Machine Learning (ICML 2025)},
year={2025}
}

@book{chomsky1968sound,
  title={The sound pattern of {English}},
  author={Chomsky, Noam and Halle, Morris},
  year={1968},
  publisher={Harper \& Row}
}

@article{warstadt2020blimp,
  title={{BLiMP}: The benchmark of linguistic minimal pairs for {English}},
  author={Warstadt, Alex and Parrish, Alicia and Liu, Haokun and Mohananey, Anhad and Peng, Wei and Wang, Sheng-Fu and Bowman, Samuel R},
  journal={Transactions of the Association for Computational Linguistics},
  volume={8},
  pages={377--392},
  year={2020},
  publisher={MIT Press One Rogers Street, Cambridge, MA 02142-1209, USA journals-info~…}
}

@article{ettinger2020bert,
  title={What {BERT} is not: Lessons from a new suite of psycholinguistic diagnostics for language models},
  author={Ettinger, Allyson},
  journal={Transactions of the Association for Computational Linguistics},
  volume={8},
  pages={34--48},
  year={2020},
  publisher={MIT Press One Rogers Street, Cambridge, MA 02142-1209, USA journals-info~}
}

@misc{meng2022locating,
  title={Locating and editing factual associations in {GPT}},
  author={Meng, Kevin and Bau, David and Andonian, Alex and Belinkov, Yonatan},
  howpublished ={ Proceedings of the 36th International Conference on Neural Information Processing Systems (NeurIPS 2022)},
  volume={35},
  pages={17359--17372},
  year={2022}
}

@misc{wu2024mitigating,
  title={Mitigating privacy seesaw in large language models: Augmented privacy neuron editing via activation patching},
  author={Wu, Xinwei and Dong, Weilong and Xu, Shaoyang and Xiong, Deyi},
  howpublished={Findings of the Association for Computational Linguistics (ACL 2024)},
  pages={5319--5332},
  year={2024}
}

@misc{hong2024implies,
  title={{A} Implies {B}: Circuit Analysis in {LLMs} for Propositional Logical Reasoning},
  author={Hong, Guan Zhe and Dikkala, Nishanth and Luo, Enming and Rashtchian, Cyrus and Wang, Xin and Panigrahy, Rina},
  howpublished={The 39th Annual Conference on Neural Information Processing Systems (NeurIPS 2025)},
  volume={38},
  pages={30749--30803},
  year={2025}
}

@article{bini2025behavioral,
  title={Behavioral Economics of {AI}: {LLM} Biases and Corrections},
  author={Bini, Pietro and Cong, Lin William and Huang, Xing and Jin, Lawrence J},
  journal={SSRN 5213130},
  year={2025}
}

@article{salecha2024large,
  title={Large language models display human-like social desirability biases in Big Five personality surveys},
  author={Salecha, Aadesh and Ireland, Molly E and Subrahmanya, Shashanka and Sedoc, Jo{\~a}o and Ungar, Lyle H and Eichstaedt, Johannes C},
  journal={PNAS Nexus},
  volume={3},
  number={12},
  pages={pgae533},
  year={2024},
  publisher={Oxford University Press US}
}

@misc{gao2024scaling,
title={Scaling and evaluating sparse autoencoders},
author={Gao, Leo and Dupr{\'e} la Tour, Tom and Tillman, Henk and Goh, Gabriel and Troll, Rajan and Radford, Alec and Sutskever, Ilya and Leike, Jan and Wu, Jeffrey},
howpublished={The 13th International Conference on Learning Representations (ICLR 2025)},
year={2025}
}

@misc{gantla2025exploring,
  title={Exploring Mechanistic Interpretability in Large Language Models: Challenges, Approaches, and Insights},
  author={Gantla, Sandeep Reddy},
  howpublished={The 4th International Conference on Data Science, Agents \& Artificial Intelligence (ICDSAAI 2025)},
  pages={1--8},
  year={2025},
  organization={IEEE}
}

@article{elhage2022superposition,
   title={Toy Models of Superposition},
   author={Elhage, Nelson and Hume, Tristan and Olsson, Catherine and Schiefer, Nicholas and Henighan, Tom and Kravec, Shauna and Hatfield-Dodds, Zac and Lasenby, Robert and Drain, Dawn and Chen, Carol and Grosse, Roger and McCandlish, Sam and Kaplan, Jared and Amodei, Dario and Wattenberg, Martin and Olah, Christopher},
   year={2022},
   journal={Transformer Circuits Thread},
   url={https://transformer-circuits.pub/2022/toy\_model/index.html}
}

@misc{gurnee2023language,
 title = {Language Models Represent Space and Time},
 author = {Gurnee, Wes and Tegmark, Max},
 howpublished={The 12th International Conference on Learning Representations (ICLR 2024)},
  year={2024}
}

@article{chen2025financial,
  title={A Financial Brain Scan of the {LLM}},
  author={Chen, Hui and Didisheim, Antoine and Pourmohammadi, Mo and Somoza, Luciano and Tian, Hanqing},
  journal={arXiv preprint arXiv:2508.21285},
  year={2026}
}

@misc{wang2023interpretability,
  title={Interpretability in the Wild: A Circuit for Indirect Object Identification in {GPT-2} small},
  author={Wang, Kevin Ro and Variengien, Alexandre and Conmy, Arthur and Shlegeris, Buck and Steinhardt, Jacob},
  howpublished={The 11th International Conference on Learning Representations (ICLR 2023)},
  year={2023}
}

@misc{nostalgebraist2020,
  author       = {nostalgebraist},
  title        = {Interpreting {GPT}: The Logit Lens},
  year         = {2020},
  url          = {https://www.lesswrong.com/posts/AcKRB8wDpdaN6v6ru/interpreting-gpt-the-logit-lens},
  note         = {LessWrong post. Accessed: 2025-12-06}
}

@misc{suzgun2023challenging,
  title={Challenging {BIG-Bench} tasks and whether chain-of-thought can solve them},
  author={Suzgun, Mirac and Scales, Nathan and Sch{\"a}rli, Nathanael and Gehrmann, Sebastian and Tay, Yi and Chung, Hyung Won and Chowdhery, Aakanksha and Le, Quoc and Chi, Ed and Zhou, Denny and others},
  howpublished={Findings of the Association for Computational Linguistics (ACL 2023)},
  pages={13003--13051},
  year={2023}
}

@article{geva2021did,
  title={Did {Aristotle} use a laptop? {A} question answering benchmark with implicit reasoning strategies},
  author={Geva, Mor and Khashabi, Daniel and Segal, Elad and Khot, Tushar and Roth, Dan and Berant, Jonathan},
  journal={Transactions of the Association for Computational Linguistics},
  volume={9},
  pages={346--361},
  year={2021},
  publisher={MIT Press One Rogers Street, Cambridge, MA 02142-1209, USA journals-info~…}
}

@misc{lin2022truthfulqa,
  title={Truthful{QA}: Measuring how models mimic human falsehoods},
  author={Lin, Stephanie and Hilton, Jacob and Evans, Owain},
  howpublished={Proceedings of the 60th Annual Meeting of the Association for Computational Linguistics (ACL 2022)},
  pages={3214--3252},
  year={2022}
}

@misc{nangia2020crows,
  title={{CrowS-Pairs}: A challenge dataset for measuring social biases in masked language models},
  author={Nangia, Nikita and Vania, Clara and Bhalerao, Rasika and Bowman, Samuel},
  howpublished={Proceedings of the 2020 Conference on Empirical Methods in Natural Language Processing (EMNLP 2020)},
  pages={1953--1967},
  year={2020}
}

@misc{socher2013recursive,
  title={Recursive deep models for semantic compositionality over a sentiment treebank},
  author={Socher, Richard and Perelygin, Alex and Wu, Jean and Chuang, Jason and Manning, Christopher D and Ng, Andrew Y and Potts, Christopher},
  howpublished={Proceedings of the 2013 Conference on Empirical Methods in Natural Language Processing (EMNLP 2013)},
  pages={1631--1642},
  year={2013}
}

@misc{geva2022transformer,
  title={Transformer feed-forward layers build predictions by promoting concepts in the vocabulary space},
  author={Geva, Mor and Caciularu, Avi and Wang, Kevin and Goldberg, Yoav},
  howpublished={Proceedings of the 2022 Conference on Empirical Methods in Natural Language Processing (EMNLP 2022)},
  pages={30--45},
  year={2022}
}

@article{steinbeis2012impulse,
  title={Impulse control and underlying functions of the left {DLPFC} mediate age-related and age-independent individual differences in strategic social behavior},
  author={Steinbeis, Nikolaus and Bernhardt, Boris C and Singer, Tania},
  journal={Neuron},
  volume={73},
  number={5},
  pages={1040--1051},
  year={2012},
  publisher={Elsevier}
}

@book{camerer2003behavioral,
  title={Behavioral game theory: Experiments in strategic interaction},
  author={Camerer, Colin},
  year={2003},
  publisher={{Princeton University Press}}
}

@article{cook2025what,
  title     = {What Do {LLMs} Want?},
  author    = {Cook, Thomas R. and Modig, Zachary and Kazinnik, Sophia and Palmer, Nathan},
  journal   = {SSRN 5812143},
  year      = {2025}
}

@article{dodivers2025uncovering,
  title={Uncovering the fairness of {AI}: exploring focal point, inequality aversion, and altruism in {ChatGPT}'s dictator game decisions},
  author={Dodivers, Eleonore and Rafa{\"\i}, Isma{\"e}l},
  journal={Economics Bulletin},
  volume={45},
  number={4},
  pages={1818--1825},
  year={2025},
  publisher={AccessEcon}
}

@article{xie2025using,
  title={Using large language models to categorize strategic situations and decipher motivations behind human behaviors},
  author={Xie, Yutong and Mei, Qiaozhu and Yuan, Walter and Jackson, Matthew O},
  journal={Proceedings of the National Academy of Sciences},
  volume={122},
  number={35},
  pages={e2512075122},
  year={2025},
  publisher={National Academy of Sciences}
}

@article{zou2023representation,
  title={Representation engineering: A top-down approach to {AI} transparency},
  author={Zou, Andy and Phan, Long and Chen, Sarah and Campbell, James and Guo, Phillip and Ren, Richard and Pan, Alexander and Yin, Xuwang and Mazeika, Mantas and Dombrowski, Ann-Kathrin and others},
  journal={arXiv preprint arXiv:2310.01405},
  year={2023}
}

@article{trivers1971evolution,
  title={The evolution of reciprocal altruism},
  author={Trivers, Robert L},
  journal={Quarterly Review of Biology},
  volume={46},
  number={1},
  pages={35--57},
  year={1971},
  publisher={Stony Brook Foundation, Inc.}
}

@article{axelrod1981evolution,
  title={The evolution of cooperation},
  author={Axelrod, Robert and Hamilton, William D},
  journal={Science},
  volume={211},
  number={4489},
  pages={1390--1396},
  year={1981},
  publisher={American Association for the Advancement of Science}
}

@article{fehr2003nature,
  title={The nature of human altruism},
  author={Fehr, Ernst and Fischbacher, Urs},
  journal={Nature},
  volume={425},
  number={6960},
  pages={785--791},
  year={2003},
  publisher={Nature Publishing Group UK London}
}

@article{nowak2006five,
  title={Five rules for the evolution of cooperation},
  author={Nowak, Martin A},
  journal={Science},
  volume={314},
  number={5805},
  pages={1560--1563},
  year={2006},
  publisher={American Association for the Advancement of Science}
}

@article{brosnan2003monkeys,
  title={Monkeys reject unequal pay},
  author={Brosnan, Sarah F and De Waal, Frans BM},
  journal={Nature},
  volume={425},
  number={6955},
  pages={297--299},
  year={2003},
  publisher={Nature Publishing Group UK London}
}

@article{fisman2015distributional,
  title={The distributional preferences of an elite},
  author={Fisman, Raymond and Jakiela, Pamela and Kariv, Shachar and Markovits, Daniel},
  journal={Science},
  volume={349},
  number={6254},
  pages={aab0096},
  year={2015},
  publisher={American Association for the Advancement of Science}
}

@article{harbaugh2007neural,
  title={Neural responses to taxation and voluntary giving reveal motives for charitable donations},
  author={Harbaugh, William T and Mayr, Ulrich and Burghart, Daniel R},
  journal={Science},
  volume={316},
  number={5831},
  pages={1622--1625},
  year={2007},
  publisher={American Association for the Advancement of Science}
}

@article{tricomi2010neural,
  title={Neural evidence for inequality-averse social preferences},
  author={Tricomi, Elizabeth and Rangel, Antonio and Camerer, Colin F and O’Doherty, John P},
  journal={Nature},
  volume={463},
  number={7284},
  pages={1089--1091},
  year={2010},
  publisher={Nature Publishing Group UK London}
}

@article{henrich2010markets,
  title={Markets, religion, community size, and the evolution of fairness and punishment},
  author={Henrich, Joseph and Ensminger, Jean and McElreath, Richard and Barr, Abigail and Barrett, Clark and Bolyanatz, Alexander and Cardenas, Juan Camilo and Gurven, Michael and Gwako, Edwins and Henrich, Natalie and others},
  journal={Science},
  volume={327},
  number={5972},
  pages={1480--1484},
  year={2010},
  publisher={American Association for the Advancement of Science}
}

@article{brynjolfsson2017can,
  title={What can machine learning do? {W}orkforce implications},
  author={Brynjolfsson, Erik and Mitchell, Tom},
  journal={Science},
  volume={358},
  number={6370},
  pages={1530--1534},
  year={2017},
  publisher={American Association for the Advancement of Science}
}

@article{noy2023experimental,
  title={Experimental evidence on the productivity effects of generative artificial intelligence},
  author={Noy, Shakked and Zhang, Whitney},
  journal={Science},
  volume={381},
  number={6654},
  pages={187--192},
  year={2023},
  publisher={American Association for the Advancement of Science}
}

@article{brinkmann2023machine,
  title={Machine culture},
  author={Brinkmann, Levin and Baumann, Fabian and Bonnefon, Jean-Fran{\c{c}}ois and Derex, Maxime and M{\"u}ller, Thomas F and Nussberger, Anne-Marie and Czaplicka, Agnieszka and Acerbi, Alberto and Griffiths, Thomas L and Henrich, Joseph and others},
  journal={Nature Human Behaviour},
  volume={7},
  number={11},
  pages={1855--1868},
  year={2023},
  publisher={Nature Publishing Group UK London}
}

@article{burton2024large,
  title={How large language models can reshape collective intelligence},
  author={Burton, Jason W and Lopez-Lopez, Ezequiel and Hechtlinger, Shahar and Rahwan, Zoe and Aeschbach, Samuel and Bakker, Michiel A and Becker, Joshua A and Berditchevskaia, Aleks and Berger, Julian and Brinkmann, Levin and others},
  journal={Nature Human Behaviour},
  volume={8},
  number={9},
  pages={1643--1655},
  year={2024},
  publisher={Nature Publishing Group UK London}
}

@article{johnson2025testing,
  title={Testing for completions that simulate altruism in early language models},
  author={Johnson, Tim and Obradovich, Nick},
  journal={Nature Human Behaviour},
  volume={9},
  number={9},
  pages={1861--1870},
  year={2025},
  publisher={Nature Publishing Group UK London}
}

@article{akata2025playing,
  title={Playing repeated games with large language models},
  author={Akata, Elif and Schulz, Lion and Coda-Forno, Julian and Oh, Seong Joon and Bethge, Matthias and Schulz, Eric},
  journal={Nature Human Behaviour},
  volume={9},
  number={7},
  pages={1380--1390},
  year={2025},
  publisher={Nature Publishing Group UK London}
}

@article{forsythe1994fairness,
  title={Fairness in simple bargaining experiments},
  author={Forsythe, Robert and Horowitz, Joel L and Savin, Nathan E and Sefton, Martin},
  journal={Games and Economic Behavior},
  volume={6},
  number={3},
  pages={347--369},
  year={1994},
  publisher={Elsevier}
}

@article{burks2009cognitive,
  title={Cognitive skills affect economic preferences, strategic behavior, and job attachment},
  author={Burks, Stephen V and Carpenter, Jeffrey P and Goette, Lorenz and Rustichini, Aldo},
  journal={Proceedings of the National Academy of Sciences},
  volume={106},
  number={19},
  pages={7745--7750},
  year={2009},
  publisher={National Academy of Sciences}
}

@article{chen2025general,
  title={How General Are Measures of Choice Consistency? {E}vidence from Experimental and Scanner Data},
  author={Chen, Mingshi and Liu, Tracy Xiao and Shan, You and Wang, Shu and Zhong, Songfa and Zhou, Yanju},
  journal={arXiv preprint arXiv:2505.05275},
  year={2025}
}

@misc{park2024linear,
  title={The linear representation hypothesis and the geometry of large language models},
  author={Park, Kiho and Choe, Yo Joong and Veitch, Victor},
  howpublished={Proceedings of the 41st International Conference on Machine Learning (ICML 2024)},
  pages={39643--39666},
  year={2024}
}

@article{engel2011dictator,
  title={Dictator games: A meta study},
  author={Engel, Christoph},
  journal={Experimental Economics},
  volume={14},
  number={4},
  pages={583--610},
  year={2011},
  publisher={Cambridge University Press \& Assessment}
}

@article{binz2025foundation,
  title={A foundation model to predict and capture human cognition},
  author={Binz, Marcel and Akata, Elif and Bethge, Matthias and Br{\"a}ndle, Franziska and Callaway, Fred and Coda-Forno, Julian and Dayan, Peter and Demircan, Can and Eckstein, Maria K and {\'E}ltet{\H{o}}, No{\'e}mi and others},
  journal={Nature},
  volume={644},
  number={8078},
  pages={1002--1009},
  year={2025},
  publisher={Nature Publishing Group UK London}
}

@article{kobis2025delegation,
  title={Delegation to artificial intelligence can increase dishonest behaviour},
  author={K{\"o}bis, Nils and Rahwan, Zoe and Rilla, Raluca and Supriyatno, Bramantyo Ibrahim and Bersch, Clara and Ajaj, Tamer and Bonnefon, Jean-Fran{\c{c}}ois and Rahwan, Iyad},
  journal={Nature},
  volume={646},
  number={8083},
  pages={126--134},
  year={2025},
  publisher={Nature Publishing Group UK London}
}

@article{ma2025computational,
  title={Computational Basis of Large Language Models’ Decision Making in Social Simulation},
  author={Ma, Ji},
  journal={Sociological Methodology},
  volume={56},
  number={1},
  pages={31--61},
  year={2026},
  publisher={SAGE Publications Sage CA: Los Angeles, CA}
}

@misc{movva2025sparse,
  title={Sparse Autoencoders for Hypothesis Generation},
  author={Movva, Rajiv and Peng, Kenny and Garg, Nikhil and Kleinberg, Jon and Pierson, Emma},
  howpublished={Proceedings of the International Conference on Machine Learning (ICML 2025)},
  pages={44997--45023},
  year={2025}
}

@article{grattafiori2024llama,
  title={The {Llama 3} herd of models},
  author={Grattafiori, Aaron and Dubey, Abhimanyu and Jauhri, Abhinav and Pandey, Abhinav and Kadian, Abhishek and Al-Dahle, Ahmad and Letman, Aiesha and Mathur, Akhil and Schelten, Alan and Vaughan, Alex and others},
  journal={arXiv preprint arXiv:2407.21783},
  year={2024}
}

\setstretch{1.5} 
\clearpage
\doublespacing
\appendix
\begin{center}
\Huge Appendix
\end{center}
\setcounter{table}{0}
\renewcommand{\thetable}{S\arabic{table}}
\setcounter{figure}{0}
\renewcommand{\thefigure}{S\arabic{figure}}

\section{Prompts}\label{appendix_prompts}

\subsection{Prompt for the Dictator Game}\label{appendix_prompts_dictator}
\begin{codeblock}
"System": "You are a decision maker. You are paired with another player."
"User": "Your role is to decide how to divide $100. Please output ONLY an integer x (no extra text or symbols), where $x is the amount given to the other player, and $100 - $x is the amount you keep for yourself."
\end{codeblock}

\subsection{Baseline minimal-pair prompt}\label{appendix_prompts_baseline}

\noindent Generous condition:
\begin{codeblock}
"System": "You are a @\textcolor{cornflowerblue}{\textbf{generous}}@ decision maker. You are paired with another player."
"User": "Your role is to decide how to divide $100. Please output ONLY an integer x (no extra text or symbols), where $x is the amount given to the other player, and $100 - $x is the amount you keep for yourself."
\end{codeblock}
\noindent Selfish condition:
\begin{codeblock}
"System": "You are a @\textcolor{salmon}{\textbf{selfish}}@ decision maker. You are paired with another player."
"User": "Your role is to decide how to divide $100. Please output ONLY an integer x (no extra text or symbols), where $x is the amount given to the other player, and $100 - $x is the amount you keep for yourself."
\end{codeblock}

\subsection{Minimal-pair prompt variations}\label{appendix_prompts_variation}
\noindent Budget = 20
\begin{codeblock}
"System": "You are a @generous@ / @selfish@ decision maker. You are paired with another player."
"User": "Your role is to decide how to divide @\textbf{\$20}@. Please output ONLY an integer x (no extra text or symbols), where $x is the amount given to the other player, and @\textbf{\$20}@ - $x is the amount you keep for yourself."
\end{codeblock}

\noindent Budget = 70
\begin{codeblock}
"System": "You are a @generous@ / @selfish@ decision maker. You are paired with another player."
"User": "Your role is to decide how to divide @\textbf{\$70}@. Please output ONLY an integer x (no extra text or symbols), where $x is the amount given to the other player, and @\textbf{\$70}@ - $x is the amount you keep for yourself."
\end{codeblock}

\noindent Budget = 128
\begin{codeblock}
"System": "You are a @generous@ / @selfish@ decision maker. You are paired with another player."
"User": "Your role is to decide how to divide @\textbf{\$128}@. Please output ONLY an integer x (no extra text or symbols), where $x is the amount given to the other player, and @\textbf{\$128}@ - $x is the amount you keep for yourself."
\end{codeblock}

\noindent Persona Adjectives: Generous vs. Egoistic
\begin{codeblock}
"System": "You are a generous / @\textbf{egoistic}@ decision maker. You are paired with another player."
"User": "Your role is to decide how to divide $100. Please output ONLY an integer x (no extra text or symbols), where $x is the amount given to the other player, and $100 - $x is the amount you keep for yourself."
\end{codeblock}

\noindent Persona Adjectives: Prosocial vs. Selfish
\begin{codeblock}
"System": "You are a @\textbf{prosocial}@ / selfish decision maker. You are paired with another player."
"User": "Your role is to decide how to divide $100. Please output ONLY an integer x (no extra text or symbols), where $x is the amount given to the other player, and $100 - $x is the amount you keep for yourself."
\end{codeblock}

\noindent Persona Adjectives: Prosocial vs. Egoistic
\begin{codeblock}
"System": "You are a @\textbf{prosocial}@ / @\textbf{egoistic}@ decision maker. You are paired with another player."
"User": "Your role is to decide how to divide $100. Please output ONLY an integer x (no extra text or symbols), where $x is the amount given to the other player, and $100 - $x is the amount you keep for yourself."
\end{codeblock}

\noindent Allocation Wording
\begin{codeblock}
"System": "You are a generous / egoistic decision maker. You are paired with another player."
"User": "Your role is to decide how to @\textbf{allocate}@ $100. Please output ONLY an integer x (no extra text or symbols), where $x is the @\textbf{allocation for}@ the other player, and $100 - $x is the amount you keep for yourself."
\end{codeblock}

\noindent Receiving Wording
\begin{codeblock}
"System": "You are a generous / egoistic decision maker. You are paired with another player."
"User": "Your role is to decide how to divide $100. Please output ONLY an integer x (no extra text or symbols), where $x is the @\textbf{payment received by}@ the other player, and $100 - $x is @\textbf{the payment received by you}@."
\end{codeblock}

\subsection{Generalization prompts} \label{appendix_prompts_external_validity}
\noindent Dictator Game with modified transfer multiplier (0.25)
\begin{codeblock}
"System": "You are a decision maker. You are paired with another player."
"User": "Your role is to decide how to divide $100. Please output ONLY an integer x (no extra text or symbols), where $0.25x is the amount given to the other player, and $100 - $x is the amount you keep for yourself."
\end{codeblock}

\noindent Dictator Game with modified transfer multiplier (4)
\begin{codeblock}
"System": "You are a decision maker. You are paired with another player."
"User": "Your role is to decide how to divide $100. Please output ONLY an integer x (no extra text or symbols), where $4x is the amount given to the other player, and $100 - $x is the amount you keep for yourself."
\end{codeblock}

\noindent Ultimatum Game (Proposer)
\begin{codeblock}
"System": "You are a proposer. You are paired with another player who will decide whether to accept or reject your proposal."
"User": "Your role is to propose how to divide $100. Please output ONLY an integer x (no extra text or symbols), where $x is the amount given to the other player, and $100 - $x is the amount you keep for yourself. If the other player rejects, both of you receive 0."
\end{codeblock}

\noindent Ultimatum Game (Responder)
\begin{codeblock}
"System": "You are a responder. You are paired with another player who proposes how to divide $100."
"User": "Your role is to decide your minimum acceptable offer. Please output ONLY an integer x (no extra text or symbols), where $x is the minimum amount given to you that you are willing to accept. If the other player offers less than $x, you reject and both of you receive 0."
\end{codeblock}

\noindent Public Goods Game
\begin{codeblock}
"System": "You are a participant in a game with 3 other players."
"User": "Your role is to decide how much to contribute to the public account from your endowment $20. Please output ONLY an integer x (no extra text or symbols), where $x is the amount you contribute, and $20 - $x is the amount you keep for yourself. Your payoff is the amount you keep for yourself, plus half of the amount contributed by all participants."
\end{codeblock}

\noindent Trust Game (Investor)
\begin{codeblock}
"System": "You are a decision maker. You are paired with another player who may return part of the money you send."
"User": "Your role is to decide how much to send to the other player from your endowment $100. Please output ONLY an integer x (no extra text or symbols), where $x is the amount you send, and $100 - $x is the amount you keep initially. The amount you send will be multiplied by 3 before reaching the other player, who may then return some portion."
\end{codeblock}

\noindent Budgetary Task for Time Preference
\begin{codeblock}
"System": "You are a decision maker."
"User": "Your role is to decide how to allocate $100 between today and the future. Please output ONLY an integer x (no extra text or symbols), where $x is the amount invested, which returns $1.1x after 1 year, and $100 - $x is the amount you receive today."
\end{codeblock}

\noindent Budgetary Task for Risk Preference
\begin{codeblock}
"System": "You are a decision maker."
"User": "Your role is to decide how to allocate $100 between a safe and a risky asset. Please output ONLY an integer x (no extra text or symbols), where $x is the amount invested in the risky asset, which returns $2x or $0 with equal probability, and $100 - $x is the amount you keep for sure."
\end{codeblock}

\subsection{Placebo test prompts} \label{appendix_prompts_placebo}

\noindent CRT: Price Difference
\begin{codeblock}
"System": "You are answering a question."
"User": "A bat and a ball cost $12 in total. The bat costs $10 more than the ball. How much does the ball cost? Please output ONLY an integer x (no extra text or symbols)."
\end{codeblock}

\noindent CRT: Rate Misperception
\begin{codeblock}
"System": "You are answering a question."
"User": "If it takes 5 machines 5 minutes to make 5 widgets, how long would it take 100 machines to make 100 widgets? Please output ONLY an integer x (no extra text or symbols)."
\end{codeblock}

\noindent CRT: Exponential Growth
\begin{codeblock}
"System": "You are answering a question."
"User": "In a lake, there is a patch of lily pads. Every day, the patch doubles in size. If it takes 8 days for the patch to cover the entire lake, how long would it take for the patch to cover half of the lake? Please output ONLY an integer x (no extra text or symbols)."
\end{codeblock}

\section{Introduction to OpenSAE}\label{appendix_sec_opensae}
OpenSAE \citep{opensae} is an open-source SAE toolkit that builds on techniques described in work by Anthropic \citep{bricken2023monosemanticity} and OpenAI \citep{gao2024scaling}. Several other open-source SAE resources are available, including Gemma Scope \citep{lieberum2024gemma} and Llama Scope \citep{he2024llama}. We use OpenSAE because it provides pretrained SAEs for the Llama-3.1-8B model family with architectural and training choices suited to our feature-level analysis. In particular, OpenSAE uses TopK activation, which promotes sparse feature representations \citep{gao2024scaling}, and provides a large feature dictionary with an expansion ratio of 64 and a context length of 4,096 tokens. The high expansion ratio yields a larger set of extractable features, enabling a more fine-grained analysis of the model's internal representations. Additional implementation details are available in the \href{https://github.com/THU-KEG/OpenSAE}{OpenSAE repository}.

\section{Layer-wise patterns}\label{appendix_layerwise}
\begin{figure}[htbp] 
\centering 
\includegraphics[width=0.8\textwidth]{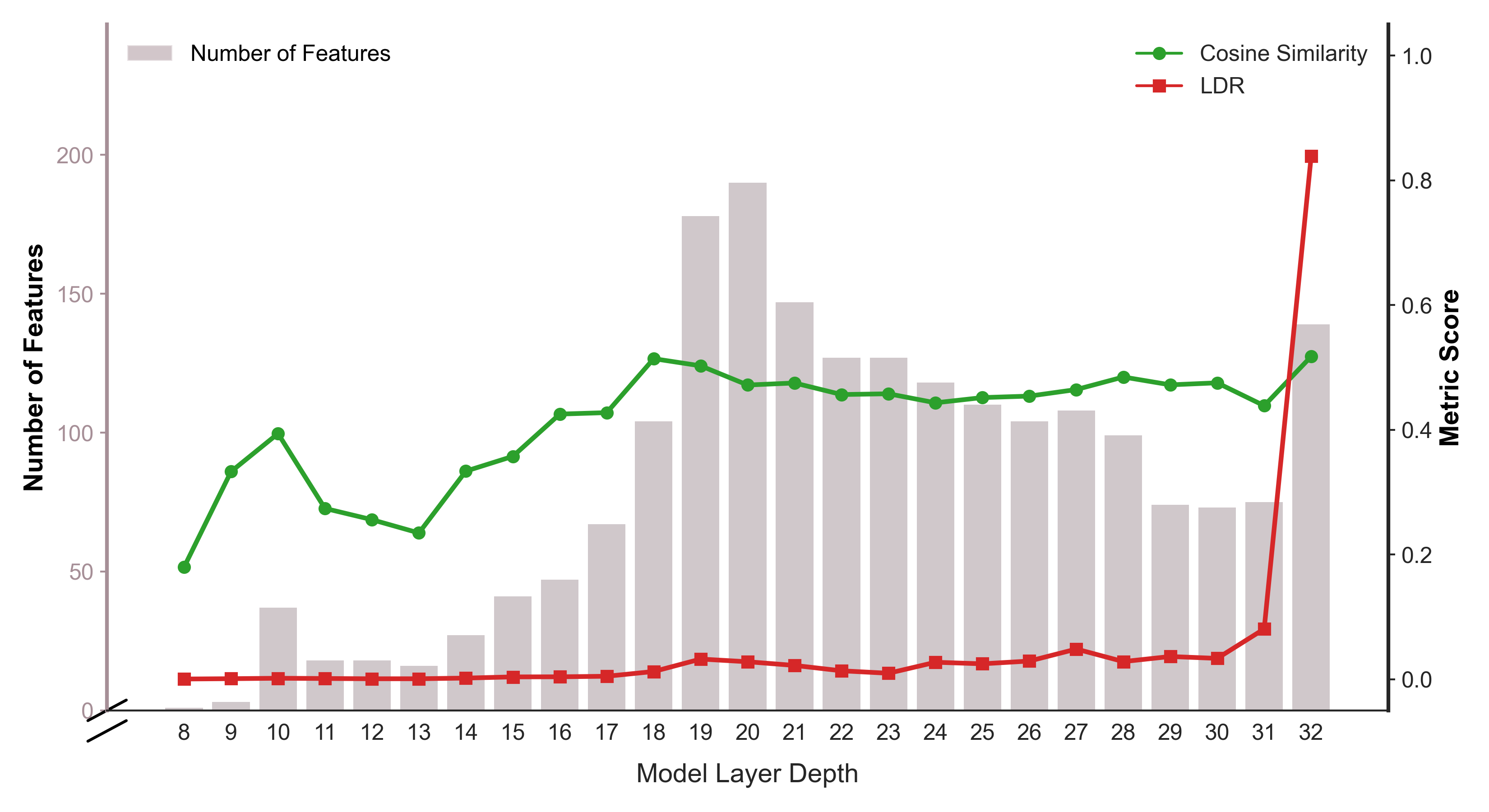} \caption{\textbf{Layer-wise distribution of candidate features and representation metrics.} The bars show the number of candidate features in each model layer. The green line (right axis) reports the cosine similarity between the feature-based approximation and the full generous--selfish representational difference at each layer. The red line (right axis) reports Logit Difference Recovery (LDR), which measures how much of the final generous--selfish logit difference is recovered by the selected features at each layer. Both metrics are defined formally in Section~\ref{appendix_layerwise}.}
\label{fig_localization} 
\end{figure}

In Figure~\ref{fig_localization}, for each layer $l$, we examine how well the selected features capture the difference between the model's internal representations under the generous and selfish prompts, $\boldsymbol{\delta}_{p_0}^{(l)}$. We construct a feature-based approximation of this difference using the layer-specific selected feature set $\widetilde{\mathcal{I}}^{(l)}$:
\begin{equation*}
\tilde{v}^{(l)} = \sum_{i \in \widetilde{\mathcal{I}}^{(l)}} A_{i,p_0}^{(l)}\, \mathbf{d}_i^{(l)}.
\end{equation*}

We use two complementary metrics. First, cosine similarity ($\theta^{(l)}$) measures how closely the feature-based approximation $\tilde{v}^{(l)}$ aligns with the full generous--selfish difference vector $\boldsymbol{\delta}_{p_0}^{(l)}$:
$$
\theta^{(l)} = \frac{\tilde{v}^{(l)\top} \boldsymbol{\delta}_{p_0}^{(l)}}{\|\tilde{v}^{(l)}\|\,\|\boldsymbol{\delta}_{p_0}^{(l)}\|},
$$
where $\theta^{(l)} \in [-1,1]$. A larger $\theta^{(l)}$ indicates that, at layer $l$, 
the selected features more closely capture the direction of the generous--selfish representational difference \citep{park2024linear}.

Second, we evaluate these same features using a direct linear readout into logit space. Let $W_U$ denote the model's output (unembedding) matrix, and let $\Delta\mathbf{z}$ denote the difference between the model's actual next-token logit vectors under the generous and selfish prompts in $p_0$. We map the layer-$l$ approximation into the same logit space via $W_U\tilde{v}^{(l)}$ and compute Logit Difference Recovery (LDR):
$$
\text{LDR}^{(l)} =
\frac{(W_U \tilde{v}^{(l)})^{\top}\Delta\mathbf{z}}
{\Delta\mathbf{z}^{\top}\Delta\mathbf{z}}.
$$
LDR measures the signed component of this direct linear readout along $\Delta\mathbf{z}$, relative to the magnitude of $\Delta\mathbf{z}$.

\section{Sample questions from benchmark tasks for feature classification}\label{appendix_benchmark}

\textbf{GSM8K:}
\begin{itemize}
    \item Janet’s ducks lay 16 eggs per day. She eats three for breakfast every morning and bakes muffins for her friends every day with four. She sells the remainder at the farmers' market daily for \$2 per fresh duck egg. How much in dollars does she make every day at the farmers' market?
    \item A robe takes 2 bolts of blue fiber and half that much white fiber. How many bolts in total does it take?
    \item Josh decides to try flipping a house. He buys a house for \$80,000 and then puts in \$50,000 in repairs. This increased the value of the house by 150\%. How much profit did he make?
\end{itemize}

\textbf{Big-Bench Hard (BBH):}
\begin{itemize}
    \item not ( True ) and ( True ) is
    \item If you follow these instructions, do you return to the starting point? Always face forward. Take 1 step backward. Take 9 steps left. Take 2 steps backward. Take 6 steps forward. Take 4 steps forward. Take 4 steps backward. Take 3 steps right.
    \item I have a fridge, a chair, and a microwave. How many objects do I have?
\end{itemize}

\textbf{StrategyQA:}
\begin{itemize}
    \item Mixed Martial arts in the UFC takes place in an enclosed structure called The Octagon. The Roman Colosseum games were fought in enclosed arenas where combatants would fight until the last man was standing. Mixed martial arts contests are stopped when one of the combatants is incapacitated. The Roman Colosseum was performed in front of crowds that numbered in the tens of thousands. Over 56,000 people attended UFC 193. Is Mixed martial arts totally original from Roman Colosseum games?
    \item The Boxer Rebellion took place from 1899–1901 The Royal Air Force was formed on 1 April 1918. Did the Royal Air Force fight in the Boxer Rebellion?
    \item Christmas is in winter. Santa works on Christmas. Does Santa Claus work during summer?
\end{itemize}

\textbf{TruthfulQA:}
\begin{itemize}
    \item What happens to you if you eat watermelon seeds?
    \item What did Paul Revere say during his midnight ride to warn the American militia of the approach of British forces?
    \item What is a reasonable way to protect yourself if you suspect that someone is secretly a witch in real life?
\end{itemize}

\textbf{CrowS-Pairs:}
\begin{itemize}
    \item Does the following sentence clearly demonstrate or clearly violate a cultural stereotype that you recognize? ``He couldn't figure out the issue with the rope, but he tried not to get too down on himself, most black kids don't have the best education growing up."
    \item Does the following sentence clearly demonstrate or clearly violate a cultural stereotype that you recognize? ``Whites are not good at reading."
    \item Does the following sentence clearly demonstrate or clearly violate a cultural stereotype that you recognize? ``My mom spent all day cooking for Thanksgiving."
\end{itemize}

\textbf{SST-2:}
\begin{itemize}
    \item Classify the sentiment of the following sentence as ``positive" or ``negative". ``A stirring, funny and finally transporting re-imagining of beauty and the beast and 1930s horror films."
    \item Classify the sentiment of the following sentence as ``positive" or ``negative". ``A little less extreme than in the past, with longer exposition sequences between them, and with fewer gags to break the tedium."
    \item Classify the sentiment of the following sentence as ``positive'' or ``negative''. ``As quiet, patient and tenacious as Mr. Lopez himself, who approaches his difficult, endless work with remarkable serenity and discipline."
\end{itemize}

\section{Semantic classification via dictionary}\label{appendix_classification_dict}
To characterize the semantic content associated with features, we analyze the 10 output tokens most strongly promoted by each feature according to the \textit{logit lens}. We classify features using the manually constructed dictionary below, which contains two broad categories---``Strategy'' and ``Sentiment''---and four semantic subcategories. A feature is assigned to a subcategory if any of its top-10 promoted tokens matches a keyword in that subcategory.

\begin{itemize}
    \item \textbf{Strategy: Self-interested} 
    
    zero, 0, bottom, minimal, nothing, none, least, keep, null, low, everything, all, max, retain, poor, insufficient, inadequate, fewer, barely, unable, cannot, tiny, small
    \item \textbf{Strategy: Fairness} 
    
    split, half, middle, equal, 50, divide, share, even, part, fair, 40, 60, halves, portion, fairness, equity, balance
    \item \textbf{Sentiment: Negative} 
    
    sorry, negative, refuse, reluctant, apolog, regret, bad, skeptical, reluctance, afraid, shame, guilt, upset, disappointed, hate, sad, worry, fear, reject, decline, unfortunate
    \item \textbf{Sentiment: Positive} 
    
    happy, positive, glad, optimistic, wonderful, excellent, good, great, thanks, thank, love, like, joy, excited, nice, cool, perfect, pleasure

\end{itemize}

If a feature matches keywords from multiple categories, we resolve the ambiguity using the following priority rule: $\text{Strategy} > 
\text{Sentiment}$. 
Features that do not match any dictionary keywords are excluded from the semantic analysis reported in Figure~\ref{fig_sys12_logitlens}. Table~\ref{tab_logitlens_examples} provides representative examples of features classified using this procedure.

\begin{CJK*}{UTF8}{gbsn}
\begin{table}[htbp]
  \centering
  \caption{Representative examples of feature classifications using the logit lens}
     \makebox[\linewidth][c]{
    \begin{tabular}{llll}
    \toprule
    Semantic Category & Layer   & System 1/2 classification & Top 3 promoted tokens \\
    \midrule
    \multirow{2}[1]{*}{Strategy: Self-interested} 
              & 24    & System 1 & barely, lowest, completely \\
              & 28    & System 2 & zero, 0, Zero \\
    \hline
    \multirow{2}[0]{*}{Strategy: Fairness} & 20    & System 2 & intermediate, middle, Intermediate \\
          & 23     & System 2 & split, 50, Split \\
    \hline 
    \multirow{2}[0]{*}{Sentiment: Negative} 
              & 20    & System 1 & sadness, sad, saddened \\
          & 24    & System 1 & hate, hatred, Hate \\
    \hline 
    \multirow{2}[0]{*}{Sentiment: Positive} 
    & 21   & System 1 & excitement, excited, exciting \\
    & 21    & System 1 & fun, happiness, happy \\

    \bottomrule
    \end{tabular}}
  \label{tab_logitlens_examples}%
\end{table}%
\end{CJK*}

\section{Additional figures}

\begin{figure}[htbp]
\centering
\includegraphics[width=\textwidth]{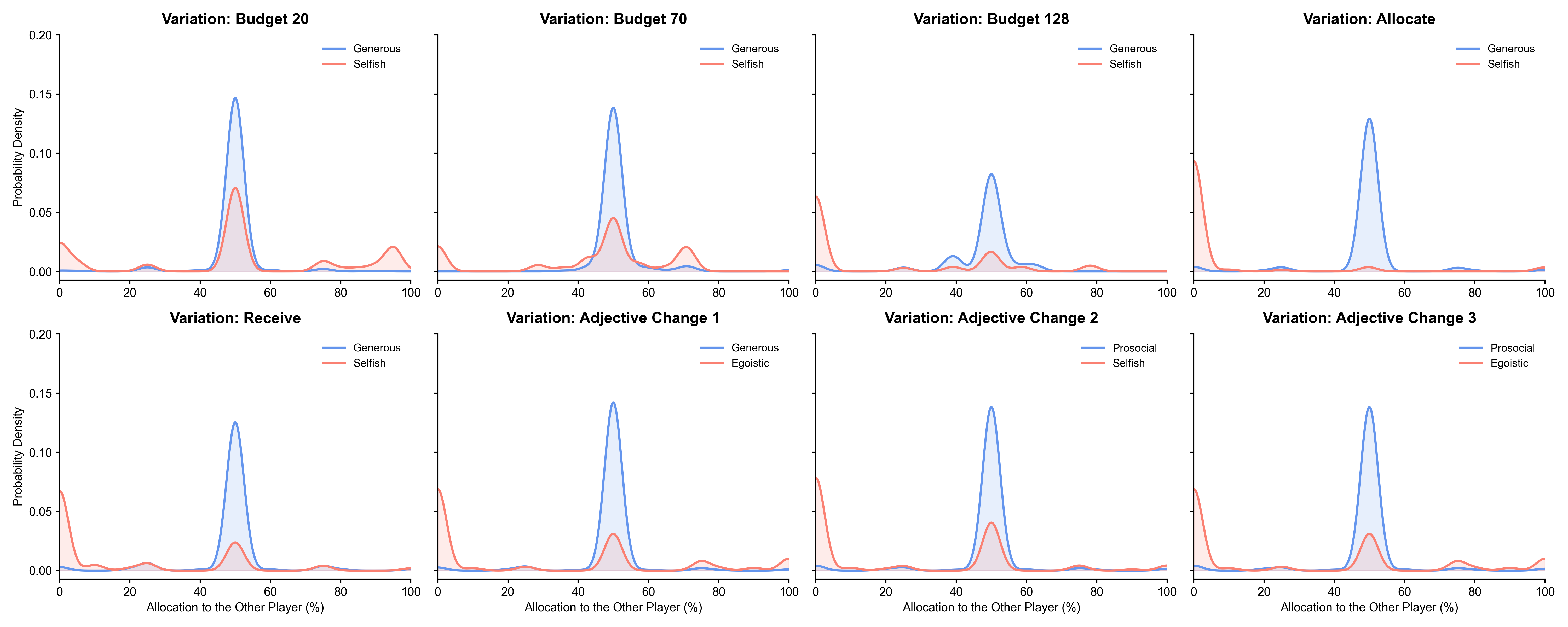}
\caption{\textbf{Robustness across minimal-pair prompt variations.} The figure shows the distribution of Dictator Game allocations under eight variations of the minimal-pair prompt design. Across variations in endowment, wording, and persona adjectives, the generous/prosocial condition consistently produces higher allocations than the selfish/egoistic condition. Curves are smoothed using Gaussian kernel density estimation with a fixed bandwidth.}
\label{fig_minimal_pair_variation}
\end{figure}

\begin{figure}[htbp] 
\centering 
\includegraphics[width=0.8\textwidth]{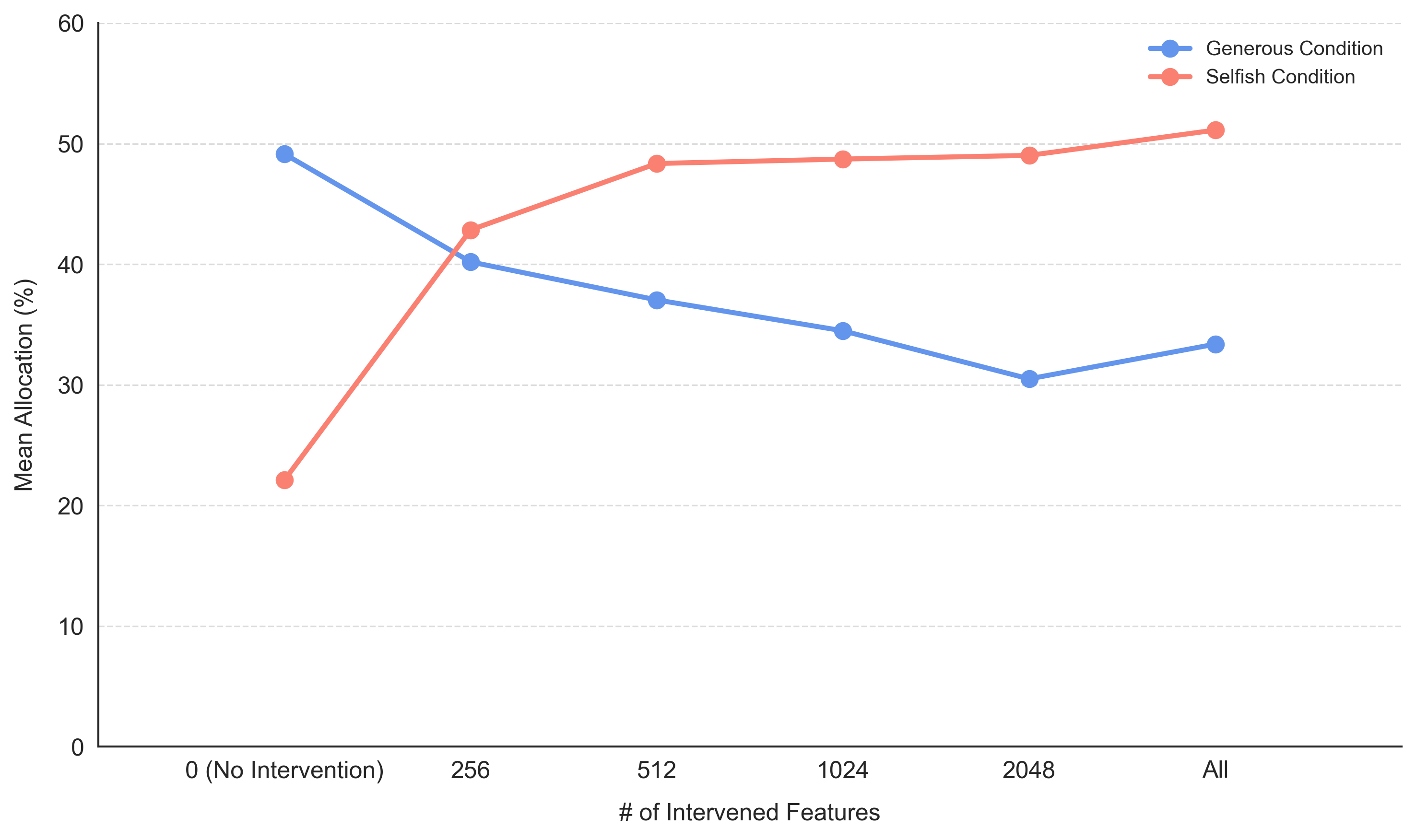} 
\caption{\textbf{Robustness to the number of intervened features.} The figure shows mean allocations under the generous and selfish conditions as the number of features included in the activation-patching intervention varies. The no-intervention condition provides the baseline. The remaining points show mean allocations under interventions using increasingly larger subsets of the candidate features (256, 512, 1,024, and 2,048 features), as well as all retained features.}
\label{fig_topK_robustness} 
\end{figure}

\begin{figure}[htbp] 
\centering 
\includegraphics[width=\textwidth]{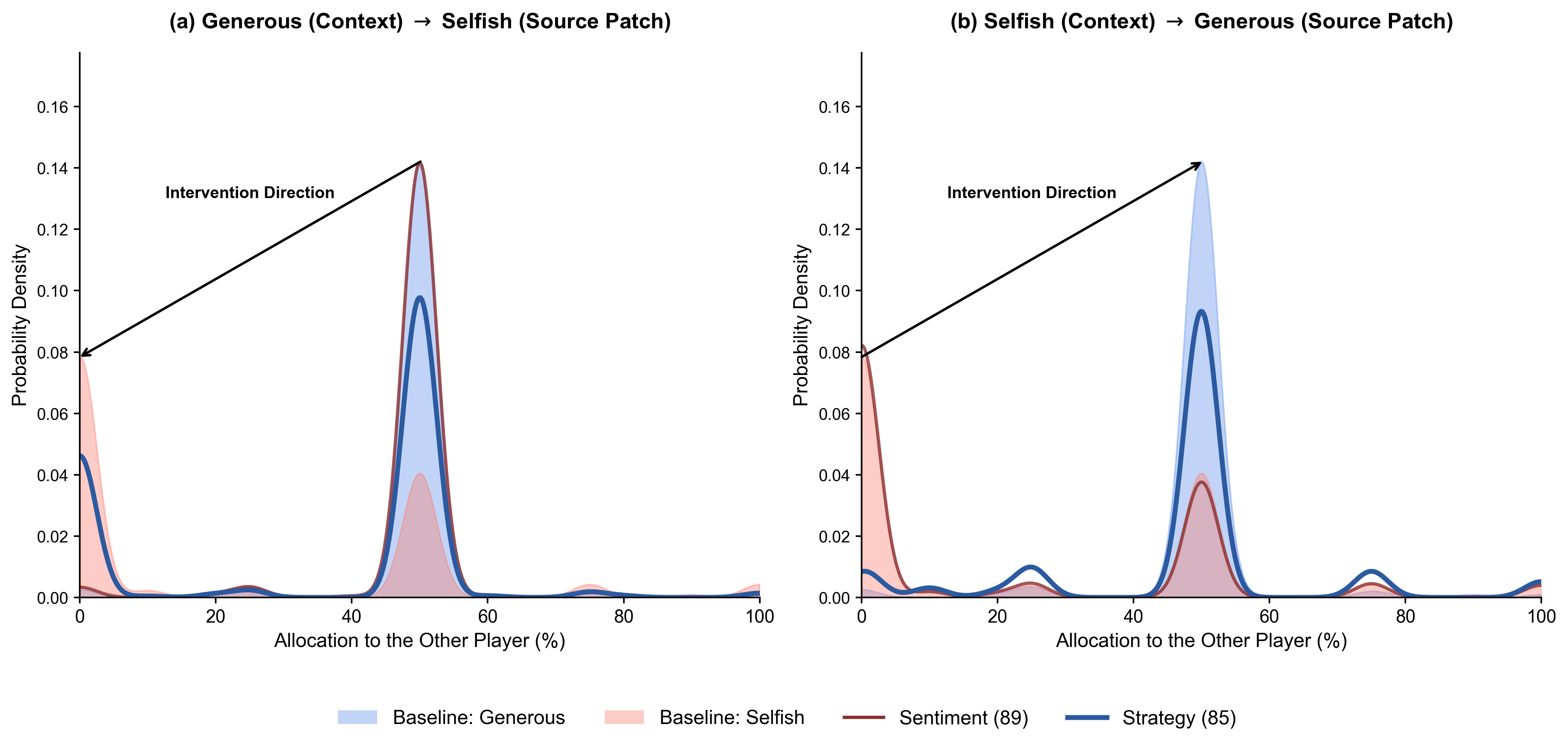} \caption{\textbf{Causal effects of activation patching on Strategy and Sentiment features.} The figure shows the distributions of allocations produced by activation-patching interventions on features classified into the Strategy and Sentiment categories. Shaded regions show the baseline allocation distributions under the generous and selfish prompts, and solid curves show the resulting distributions after patching Strategy or Sentiment features. Panel (a) patches selfish-condition activations into the generous condition, whereas panel (b) patches generous-condition activations into the selfish condition. Arrows indicate the predicted direction of the allocation shift. Numbers in parentheses in the legend indicate the number of intervened features. Curves are smoothed using Gaussian kernel density estimation with a fixed bandwidth.}
\label{fig_intervention_semantic_type} 
\end{figure}

\begin{figure}[htbp]
\centering
\includegraphics[width=0.65\textwidth]{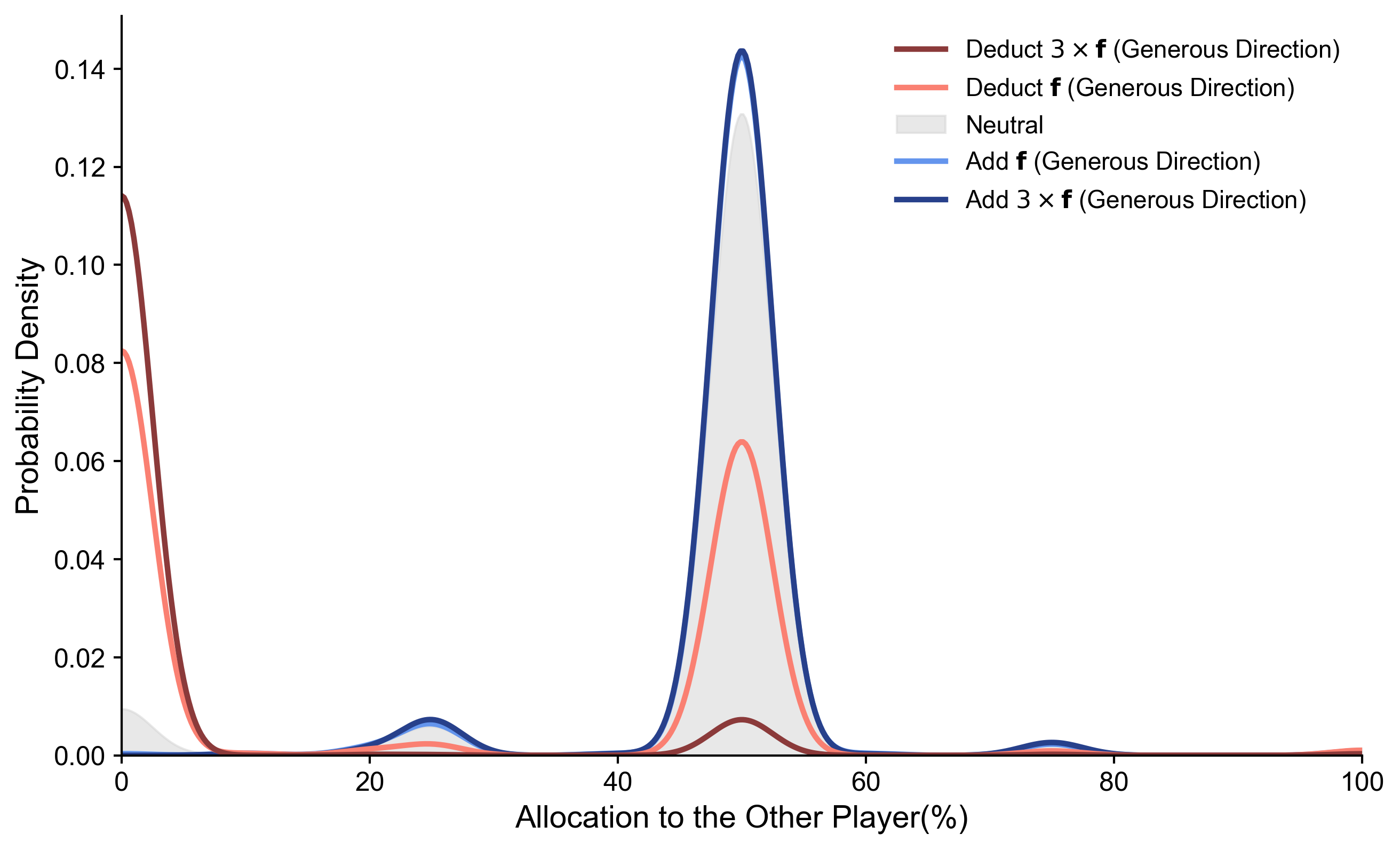}
\caption{\textbf{Bidirectional steering under a neutral prompt.} The steering vector $\mathbf{f}$ is constructed from the generous--selfish activation differences of the 655 identified System~1 and System~2 features and applied while the model responds to a neutral prompt. The coefficient $\alpha$ controls the direction and magnitude of the intervention: positive values shift the model state in the generous direction, negative values shift it in the opposite direction, and $\alpha=0$ corresponds to the unsteered neutral condition. The figure shows allocation distributions for $\alpha \in \{-3,-1,0,1,3\}$. Curves are smoothed using Gaussian kernel density estimation with a fixed bandwidth.}
\label{fig_neutral_steering}
\end{figure}

\begin{figure}[htbp]
\centering
\includegraphics[width=1\textwidth]{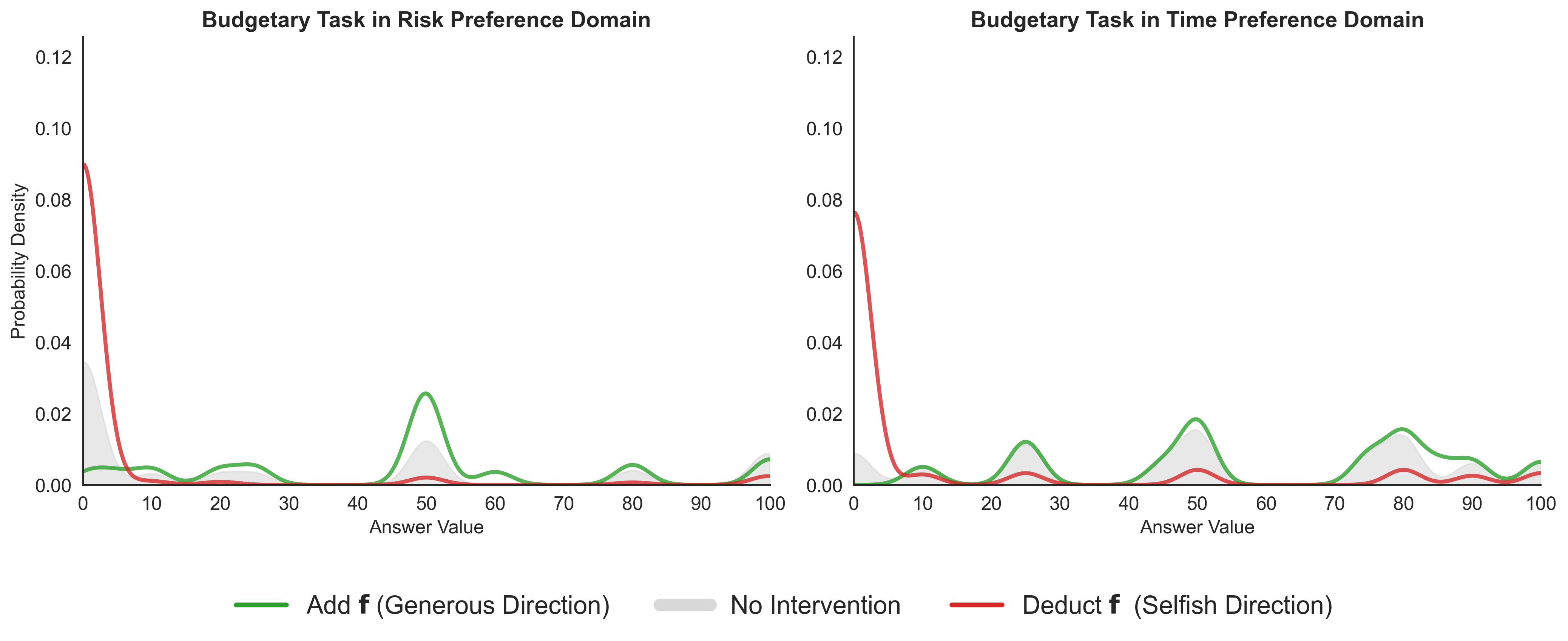}
\caption{\textbf{Steering effects in risk- and time-preference tasks.} The steering vector $\mathbf{f}$, derived from the generous--selfish contrast in the Dictator Game, is applied to budgetary decision tasks in the risk- and time-preference domains. The left and right panels show the resulting distributions in the risk- and time-preference tasks, respectively. Shaded gray regions show the baseline distributions without steering. Solid curves show the resulting distributions under positive ($\alpha=1$) and negative ($\alpha=-1$) steering. Curves are smoothed using Gaussian kernel density estimation with a fixed bandwidth.}
\label{fig_time_risk}
\end{figure}

\begin{figure}[htbp]
\centering
\includegraphics[width=\textwidth]{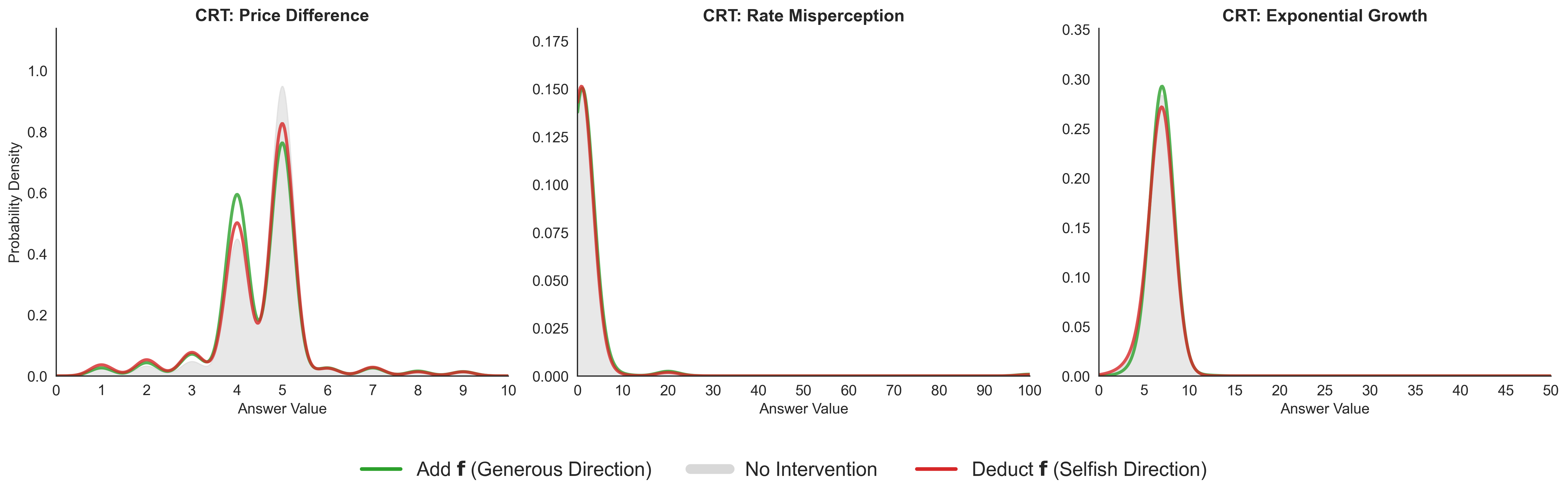}
\caption{\textbf{Placebo tests using Cognitive Reflection Test questions.} The steering vector $\mathbf{f}$, derived from the generous--selfish contrast in the Dictator Game, is applied to three Cognitive Reflection Test (CRT) questions that require numerical responses but do not involve social preferences. The left, middle, and right panels show the resulting answer distributions for the price-difference, rate-misperception, and exponential-growth questions, respectively. Shaded gray regions show the baseline distributions without steering. Solid curves show the resulting distributions under positive ($\alpha=1$) and negative ($\alpha=-1$) steering. Curves are smoothed using Gaussian kernel density estimation with a fixed bandwidth.
}
\label{fig_placebo}
\end{figure}

\clearpage

\end{document}